\documentclass{aa}

\usepackage[varg]{txfonts}
\usepackage{graphicx}
\usepackage{array, booktabs, makecell}
\usepackage{siunitx}
\usepackage[version=4]{mhchem}
\usepackage{textcomp, gensymb}
\usepackage{amsmath}
\usepackage{amsfonts}
\usepackage{comment}
\usepackage{multicol}
\usepackage[utf8]{inputenc}
\usepackage[english]{babel}
\usepackage{natbib}
\bibpunct{(}{)}{;}{a}{}{,}
\usepackage[hidelinks,colorlinks=true,linkcolor=blue,citecolor=blue]{hyperref}
\usepackage[dvipsnames]{xcolor}

\newcommand{\be}{\begin{equation}}
\newcommand{\ee}{\end{equation}}
\newcommand{\bea}{\begin{eqnarray}}
\newcommand{\eea}{\end{eqnarray}}

\begin{document}

\title{AGN hosting jets. II: Deriving fundamental information from semi-analytical models and the LoTSS sample}

\author{Pau Beltr\'an-Palau \inst{1}, Manel Perucho \inst{1,2}, Jos\'e-Mar\'{\i}a Mart\'{\i}\inst{1,2}, Xavier L\'opez-L\'opez \inst{3}}

\institute{
\inst{1} Departament d'Astronomia i Astrof\'{\i}sica, Universitat de Val\`encia, Av. Vicent Andr\'es Estell\'es 19, 46100, Burjassot, Valencian Country, Spain \\
\inst{2} Observatori Astron\`omic, Universitat de Val\`encia, C/ Catedr\`atic Jos\'e Beltr\'an 2, 46980, Paterna, Valencian Country, Spain \\
\inst{3} INAF-Osservatorio di Astrofisica e Scienza dello Spazio di Bologna, Via Piero Gobetti 93/3, 40129 Bologna, Italy
}

\date{Received xx / Accepted yy}

  \abstract
   {Galactic activity has been shown to play a fundamental role in galaxy and cluster evolution throughout the history of the Universe. In the last decades, a number of works have tackled the determination of radio luminosity functions and their evolution, with the aim to try to understand and quantify this role. In this work we propose a different approach, based on the application of a semi-analytical model to produce mock populations of radio loud active galaxies throughout several orders of magnitude in power, and the comparison with observational samples.}
{The aim of this work is to test the ability of our semi-analytical model to reproduce luminosity and size distributions of large numbers of sources from the radio-loud active galactic nuclei (RLAGN) sample from the LoTSS survey. This is an initial step in the potential application of the method 'to probe high-redshift populations and eventually study the evolution of galactic activity via comparison with current and future surveys to validate our approach.}
{We apply our semi-analytical model for the evolution of radio galaxies and generate hundreds of thousands of sources applying random selection of parameters within probability distributions for each of them. Then we compare our results with those of RLAGN in the LoTSS sample.}
{Our results show that the model is able to reproduce the main features of the observed sample, and identifies jet mass-load included in our model and activity periods as critical parameters that should be addressed in more detail in future work. These initial results allow us to establish a maximum activity period ranging from 10 to 200~Myr within a uniform logarithmic distribution, i.e., with an active time distribution probability $p(t_{\rm max})\propto t_{\max}^{-1}$. We also reproduce the radio luminosity function for the $z<0.6$ population, and the jet power spectrum from $10^{34.5}\,{\rm W}$ to $10^{38}\,{\rm W}$. Our results show that mass-load is a fundamental parameter in the determination of radio source luminosities with distance, mainly in the low-power regime. Here we use a characterization of mass-load that accounts for dependence on the jet power and galaxy size. Our model predicts that at least $3-4\times 10^{5}$ radio galaxies must exist in the LoTSS field in order to obtain the 17045 observed ones. This population is vastly dominated by low power sources.}
   {}


\titlerunning{Fundamental information from semi-analytical models and the LoTSS sample}
\authorrunning{Beltr\'an-Palau, Perucho, Mart\'{\i} \& L\'opez-L\'opez}

\maketitle

\section{Introduction}

Active galactic nuclei (AGN) are known to play a fundamental role in galaxy and cluster evolution \citep[][]{2007ARA&A..45..117M,2012ARA&A..50..455F}. The impact of nuclear activity occurs mainly through the evolution of jets within the host galaxy and cluster, after they are formed by the extraction of rotational energy from supermassive black holes (SMBH) at the active nucleus \citep{1977MNRAS.179..433B}. It seems clear that jet propagation indeed plays a significant role in the evolution of its host galaxy and its environment. In particular, galactic activity has been shown to be crucial to prevent galactic halo cooling both by semi-analytical models \citep[][]{2006MNRAS.365...11C,2006MNRAS.370..645B} and numerical simulations \citep[e.g.,][]{2020NatRP...2...42V,2023MNRAS.523.1104W,2026A&A...706L..13W}, through repeated energy injection on cluster scales, as also confirmed by X-ray observations \citep[][]{1993MNRAS.264L..25B,2000MNRAS.318L..65F,2005ApJ...635..894F}. However, to which extent this mechanism has been principal along the history of the Universe is still unknown. Different works have tackled the relevance of galactic activity through cosmic time \citep[e.g.,][]{1990MNRAS.247...19D,1999MNRAS.302..515K,1999MNRAS.304..160J,2001MNRAS.322..536W,2004MNRAS.355L...9R,2007MNRAS.381.1548K,2023MNRAS.523.5292K}, focusing mainly on the study of the radio luminosity function (RLF) of AGN. 

\cite{1990MNRAS.247...19D} report on a redshift cut-off of the AGN number between $z=2$ and 4, concluding that both flat (compact, e.g., quasars), and steep spectrum (extended, lobe dominated, e.g., radio galaxies) sources undergo the same cosmological evolution. Their results point towards pure luminosity or luminosity/density evolution as the main drivers of the RLF evolution, with the high luminosity population showing stronger changes. This approach was later supported by \cite{2001MNRAS.328..897S}. In contrast, \cite{1999MNRAS.302..515K} advocate for a density evolution dominating the RLF changes with redshift. 

\citet{1999MNRAS.304..160J} propose two populations, associated to FRI/FRII morphologies \citep{1974MNRAS.167P..31F}, where the FRI population is independent of redshift, whereas powerful FRIIs with strong emission lines would show significant cosmological evolution. However, the authors also suggest that the two populations could be related to different accretion types (namely high-excitation, HEGs, and low excitation, LEGs, galaxies) and not strictly to FRIs and FRIIs. Additionally, they include beaming considerations to study 5-GHz populations and attain agreement with observational counts within the AGN unification scenario. This result is also supported by \cite{2001MNRAS.322..536W}. In this context, \citet{2007MNRAS.381.1548K} showed that the RLF broken distribution can be explained within the FRI/FRII dichotomy, with the former dominating below the break and the latter above it. \citet{2012MNRAS.421.1569B} showed that LEGs represent the majority in the low luminosity region, whereas both LEGs and HEGs populate the high luminosity fraction above the break.

More recent works \citep[see, e.g.,][]{2023MNRAS.523.5292K} have studied the cosmic evolution of LEGs \citep[see also][]{2013MNRAS.430.3086G} using a subset of 11783 AGN galaxies from the LoTSS sample. The authors confirmed that LEGs dominate the AGN population in the local Universe ($z \leq 0.75 - 1.0$), supporting the idea that cooling hot gas from galactic haloes feeds those nuclei. 

From an observational perspective, different telescopes, such as the X-ray telescope Swift/XRT \citep[e.g.,][]{2022A&A...663A.147S,2022ApJ...940...77M}, the infrared telescope James Webb \citep[e.g.,][]{2022ApJ...939L..28D,2025Natur.639..897W,2025NatAs.tmp..191S}, and even multi-wavelength campaigns \citep[e.g.,][]{2025NatAs...9..293B}, have opened the way to the study of quasars at high redshifts, showing that the $z=2$ AGN golden era could be the consequence of an observational limitation and that galactic activity could have been crucial since very early in the history of the Universe. \citet{2022A&A...663A.147S} pointed out that radio-loud AGN might have been more abundant at $z=4$ and that jets may have thus played a central role in the fast formation of SMBHs at those early times via the extraction of accretion disc angular momentum. If powerful jets were formed in those rapidly accreting SMBHs, the feedback driven by the jet on the ambient is a relevant aspect of galaxy formation and evolution that needs to be addressed. 

These facts clearly indicate the relevance of understanding AGN jet cosmic evolution via population statistics and luminosity functions at different wavelengths. Actually, cosmological simulations incorporate galactic activity via energy injections in accreting galaxies \citep[e.g.,][and references therein]{2020NatRP...2...42V}. However, the inclusion of galactic activity in these simulations does not take into account the nature of jets as relativistic and collimated outflows. The consistent inclusion of AGN evolution and impact in cosmological simulations is still a challenge in the current state-of-the-art, which could be solved by applying semi-analytical models. 

Another unknown, critical parameter is the duration of activity, for which we can only place limits using population studies. \citet{2020MNRAS.496.1706S} used the RAiSE evolution model \citep{2015ApJ...806...59T} to study LOFAR samples and concluded that the maximum activity time is $t_{\rm on}\propto t^{-1}$ and that the distribution of jet power is $P(L_0)\propto L_0^{-1}$. More recently, \citet{2025MNRAS.537..343Q} used a sample of 79 remnant sources in the local Universe ($z<0.2$) and confirmed this result for the activity time. They also obtained a jet power distribution $P(L_0)\propto L_0^{-1.5}$ for sources with powers between $10^{36.5}$ and $10^{40}$~W. 

In this paper, we take an approach based on the generation of a mock population of galaxies using a semi-analytical model of radio galaxy (RG) evolution. This approach has been followed in the past by different authors to study active galaxy populations from different surveys and their corresponding contexts \citep[e.g.,][]{2006MNRAS.372..381B,2008ApJ...682L..17B,2007MNRAS.381.1548K,2015ApJ...806...59T,2019A&A...622A..12H}. In \citet{2006PhDT........38B,2006MNRAS.372..381B,2007ApJ...658..217B,2008ApJ...682L..17B}, the authors used dynamical and radiative models to study the properties and density of a complete population of RGs giving rise to the 3CRR, 6CE, and 7CRS sources. \citet{2006MNRAS.372..381B,2007ApJ...658..217B}
used a Montecarlo simulation on jet power, keeping the host galaxy properties fixed in all the simulated radio sources. This study was limited to the powerful sources catalogued in those samples. \citet{2007MNRAS.381.1548K} followed a similar approach, using the evolution model presented in \citet{1997MNRAS.286..215K} and the radiative calculations proposed in \citet{1997MNRAS.292..723K} to produce separate RLFs for FRI and FRII sources and make theoretical predictions on the cosmological evolution of these populations. 
 
In \citet{2025A&A...704A.320B} (Paper~I from now on) we presented a new model, which, although similar to those published by \citet{2015ApJ...806...59T} and \citet{2018MNRAS.475.2768H}, is simpler than the first and, with respect to the second, introduces changes in the evolution of low-power jets (i.e., those typically hosted in low-luminosity jetted AGN). The model introduces mass-loading into the jet flow and allows us to study the evolution of RGs within a wide range of jet powers. In this paper, we use that model to generate a population of radio sources to be compared with observed samples. We will focus on the LoTSS sample up to redshift $z=0.6$ to test its predicting power before extending our study to larger redshifts. Our strategy is to calibrate the model using the low redshift population, where we can use the most complete sample possible. Once this calibration is done, we can use the model to make predictions for the high redshift population by trying, e.g., different RLF evolution models.

Our simulated sample is generated by modifying the fundamental parameters (within plausible values) that determine jet evolution and their observed radio luminosities, using either randomized uniform or appropriate probability distributions for each of them. These include power at injection, activity periods, host galaxy properties, viewing angles, or age at observation time.

The paper is organized as follows. In Sect.~\ref{method_description} we present the description of the method followed to run a simulation and create a population of RGs, together with the probability distributions used for each of the relevant parameters. 
We present the results of the simulation in Sect.~\ref{results} and compare them to the RG distribution in the LoTSS sample. In Sect.~\ref{sec:disc}, we discuss the limitations of this approach and the possible ways in which we can improve it in future work. Finally, we present a summary and the conclusions of this work in Sect.~\ref{conclusions}. The cosmological parameters used in this paper are $H_0=70\,\text{km s}^{-1}$, $\Omega_R=0$, $\Omega_M=0.3$ and $\Omega_\Delta=0.7$. 

\section{Method description}\label{method_description}

Using the evolutionary model for RGs presented in Paper~I, we generate a mock population of radio-loud AGNs (hereafter RLAGN) to be compared with the corresponding  RLAGN sample in the LoTSS survey \citep{2019A&A...622A..12H}. This is achieved by randomizing some of the required parameters, according to appropriate probability distributions, while keeping others fixed. We limit our study to sources with redshifts $z\leq 0.6$ and use the redshift and luminosity distributions for those RLAGN within the LoTSS sample as a constraint for our simulation. The simulation is then expected to generate a jet power spectrum that reproduces the observed luminosity distribution and check whether it also fits the source size distribution, within a reasonable range of host-galaxy and jet parameters.

Once all the necessary parameters are determined, we can then apply the evolution model. At the randomly selected observing age of the RG, we record its projected length on the plane of the sky, the radio luminosity, the received flux at $150\,\text{MHz}$, and the spectral index. We calculate the luminosities at two frequencies: $ \nu = 150 \, \text{MHz} $ (observation frequency) and $ \tilde{\nu} = (1+z)\nu $ (emission frequency). The first one ($L_\nu$) will be compared with the LoTSS sample (which gives the luminosities at rest frame 150 MHz) and the second one ($L_{\tilde\nu}$) will be used to calculate the observed flux and estimate the radio-lobe spectral index (together with $L_\nu$).

To reduce the computing time (since we will have to calculate approximately $10^6$ luminosities in each simulation), we apply a numerical approximation for the luminosity integral (Eq.~28 in Paper~I). In particular we apply the  composite Simpson's 1/3 rule at order 100. As a test, we calculated the ratio between the exact and the approximated value for 1000 luminosities with random initial parameters. We obtained a mean of 0.9, and a standard deviation of 0.3. We find this to be acceptable given the approximations made by our model and the uncertainties in the characterization of RGs. 

The radio-lobe spectral index $ \alpha $ is derived by using the relation $ L_\nu = L_{\tilde\nu}(1+z)^\alpha $. Finally, we compute the observed flux from:
\be\label{flux}
S_\nu=\frac{L_{\tilde\nu}}{4\pi r(z)^2(1+z)},
\ee
where $r(z)$ is the coordinate defined by the direction to the observer, given by 
\be
r(z)=\frac{c}{H_0}\int_0^z \left(\Omega_R(1+z')^4+\Omega_M(1+z')^3+\Omega_\Delta\right)^{-1/2}dz',
\ee
with $z'$ the integral variable. We use the derived flux for each source at the observing time  to determine whether a radio galaxy would be observable by LOFAR, taking into account that its flux detection limit is $ S_\nu = 0.5 \, \text{mJy}$. However, as shown in \citet{2019A&A...622A..12H}, the LoTSS sample is also limited by surface brightness, with a threshold of 100~$\mu$Jy per beam. We apply this threshold by computing the projected angular surface of the source on the plane of the sky and dividing the total flux by this surface expressed in number of beams (assumed to be circular with a radius of 6 arcseconds), where the surface of the source is calculated as that of an ellipse with major axis corresponding to the jet propagation direction and minor axis the computed radial size. 

Once we have generated our simulated sample, we compare it with the observations. Within the LoTSS RLAGN sample, we only consider sources with $z \leq 0.6$, to avoid the region at which a strong drop in the source density takes place, caused by the increasing lack of redshift measures, among other reasons. From the public data, we estimate the projected lengths of the unresolved sources (those with angular sizes smaller than 1 arcsec) in the same way as in \cite{2019A&A...622A..12H}, i.e., taking the measured deconvolved major axis plus three times the error on this axis as an upper limit of the size. We then exclude the unresolved sources that do not have an estimate of the angular size (65 sources). This limits the RLAGN sample to a total of 17045 sources, with 996 of them (5.8\% of the total) classified by \cite{2019A&A...622A..12H} as quasars. We merge both populations together in our simulated set because we cannot apply any optical classification. We remind the reader that our model does not include jet emission, so that we implicitly assume that the radio emission at low frequencies is dominated by the lobes. The assumption is supported by evidence that the radio lobes in powerful quasars can show a brightness of the order of that emitted by the jet at these frequencies, even at high redshifts \citep[see, e.g., the case of 0836+710 at $z=2.2$; ][]{2019A&A...631A..49K}. Therefore, although this assumption underestimates the emission of aligned sources, we do not expect this to significantly affect the detectability of more than a very minor portion of our sample.

Compact sources are known to be affected by self-synchrotron and free-free absorption within the host galaxy. For this reason, the number of observed sources is expected to be smaller than that we obtained by not taking absorption into account. Our model does not include this effect, so any comparison between the simulated and observed populations can be plainly wrong, starting from the use of the received flux to assign a luminosity to a source. Nevertheless, we keep compact sources in the model and in our plots and account their numbers as a lower limit of the intrinsic, complete population. We have evaluated the impact of this effect and show that it mainly affects the low-power population, which is precisely the most numerous one, too.

In the next subsections we give a detailed explanation of the parameters required to do the simulation and how are they assigned. Table~\ref{table_param} gives a summary of all this information. The selected values and intervals of randomization are inspired by the results of Paper I.

\subsection{Parameters that define the host galaxies and clusters} \label{ambient}
To generate the simulated population, we first determine the initial parameters of each of their host galaxies. As explained in Paper~I, we model the ambient density profile with a spherically symmetric double King profile. The host galaxies are thus characterized by a core radius and density ($a_g$ and $\rho_{g,0}$, respectively), a cluster (or group) core radius and density ($a_c$ and $\rho_{c,0}$ respectively), the exponents that fix the density profiles ($\beta_g$ and $\beta_c$ for the galaxy and cluster, respectively), and an inter-cluster density, $\rho_{\rm ICM}$: 
\be\label{density}
\rho(l)=\rho_{g,0}\left(1+\left(\frac{l}{a_g}\right)^2\right)^{\beta_g/2}+\rho_{c,0}\left(1+\left(\frac{l}{a_c}\right)^2\right)^{\beta_c/2}+\rho_{\rm ICM},
\ee
where $l$ is the radial coordinate. 

The relevant parameters that determine the profile are randomly chosen from a uniform probability distribution within the following intervals, inspired by the results of Paper~I: $a_{g} \in [0.5, 2] \text{\ kpc}$, $a_{c} \in [36, 60] \text{\ kpc}$, $\beta_g \in [-2, -1.6]$, $\beta_c$ $\in [-1.5, -1.1]$ and $\rho_{g,0} \in [0.1, 10] \text{ protons cm}^{-3}$. The minimum value of $a_{c}$, the radius of the cluster, is set to $36$~kpc to ensure that, for any combination of parameters, the resulting density profile has the shape of a double King profile. The cluster core density is fixed depending on the galactic one as $\rho_{c,0}=\rho_{g,0}/100$. The ICM density is taken as redshift dependent to account for the cosmological expansion \citep[e.g.,][]{2018A&A...617L...3B}: $\rho_{\rm ICM} =(3.345 \times 10^{-28} \mathrm{~kg} / \mathrm{m}^3)(1+z)^3$. We have not included any redshift dependence in the rest of the ambient medium parameters because we limit our comparison with the observed sample to redshifts up to $0.6$.

The spherical gas distribution locates our galaxies at the symmetry center of groups or clusters. However, only the most powerful ones reach distances beyond the group/cluster core, as we will see later. Therefore, the requirement of spherical symmetry is limited de facto to the distance to which the jet propagates until the observing time, i.e., a smaller region than the much larger pre-designed ambient medium. This favours the validity of the assumption in the case of low power sources, which typically propagate up to smaller distances and represent a vast majority. 

\begin{figure}[htbp]
\begin{center}
\includegraphics[width=90mm]{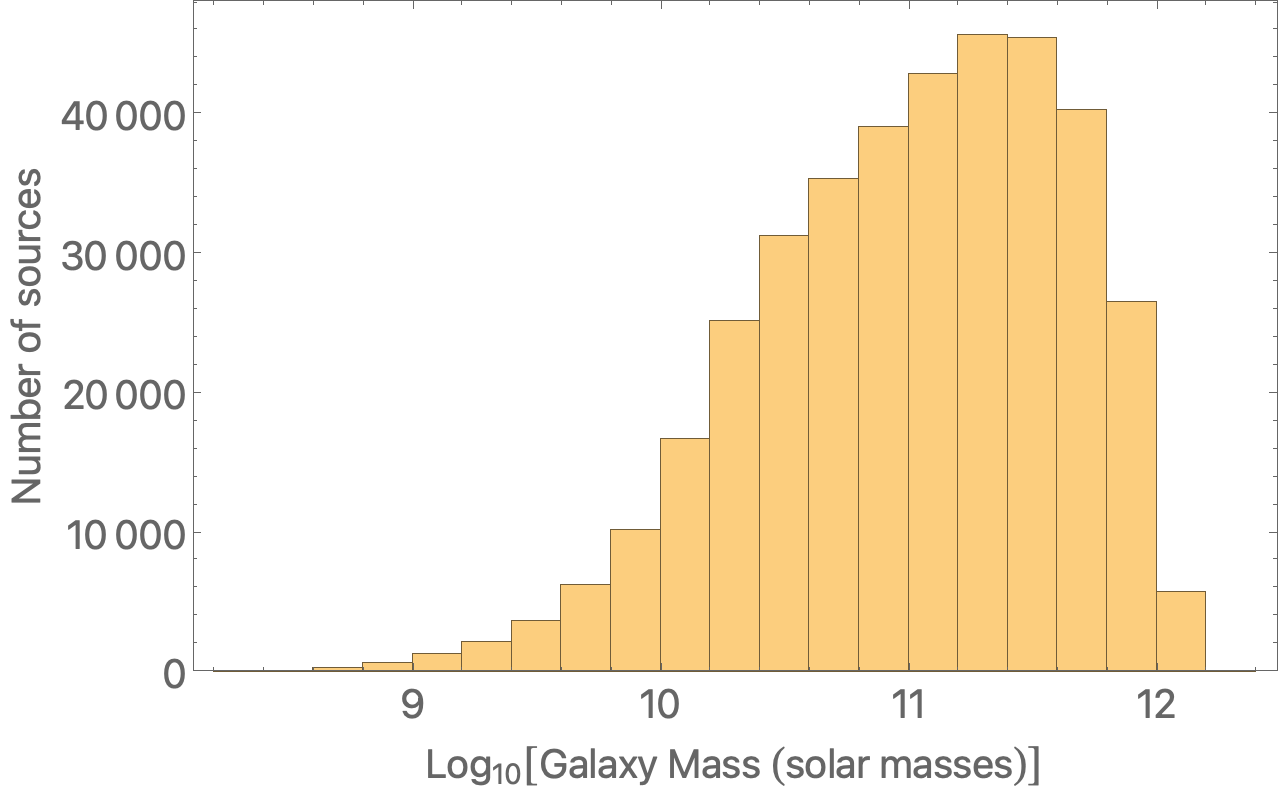}
\includegraphics[width=90mm]{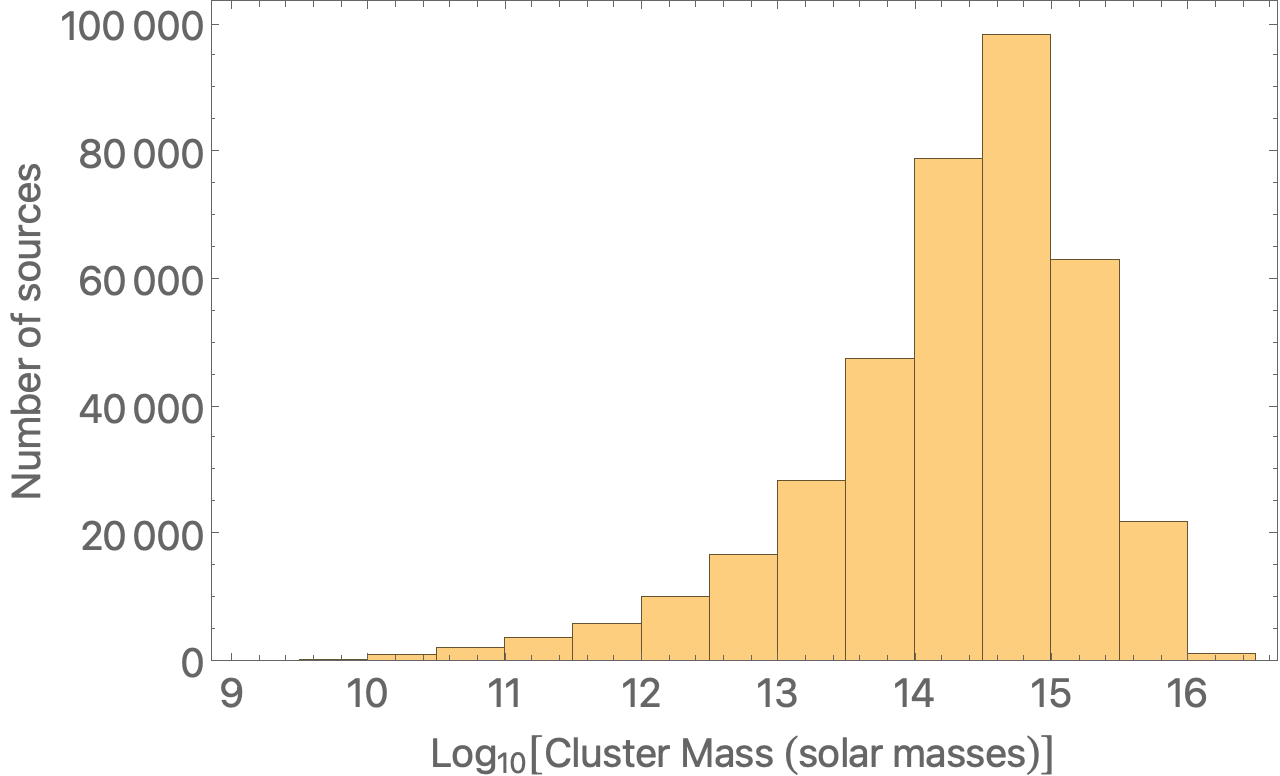}
\end{center}
\caption{The top panel shows a histogram of the galaxy mass distribution, estimated as the enclosed gas up to 20 times the galactic core radius, which results in galactic sizes between 10 and 40~kpc. The bottom panel shows the distribution of the group/cluster mass estimated as the enclosed mass up to $\rho_{200}$ (the density that is 200 times larger than the critical density of the Universe, and considering that the host galaxy lies at the centre of the cluster).}
\label{fig:masses}
\end{figure}

The uniform probability distribution of central densities produces a bias towards larger masses. As a consequence, this generates a mass distribution that is also shifted towards high values, in agreement with the association of radio-loud sources with massive ellipticals \citep[e.g.,][]{2005MNRAS.362...25B,2019A&A...622A..17S}. The resulting galaxy/cluster mass distribution is shown in Fig.~\ref{fig:masses}. The bottom panel shows the group/cluster masses assuming that the host galaxy is at the centre of the distribution.

The adiabatic index of the ambient medium is taken to be $\Gamma=5/3$, and the sound speed is fixed to $c_{\rm s}=6.84 \times 10^2\,\text{km/s}$ (corresponding to an ambient temperature of $T_{\rm a} = 1.7\times 10^7$ K).

\begin{figure}[htbp]
\begin{center}
\includegraphics[width=90mm]{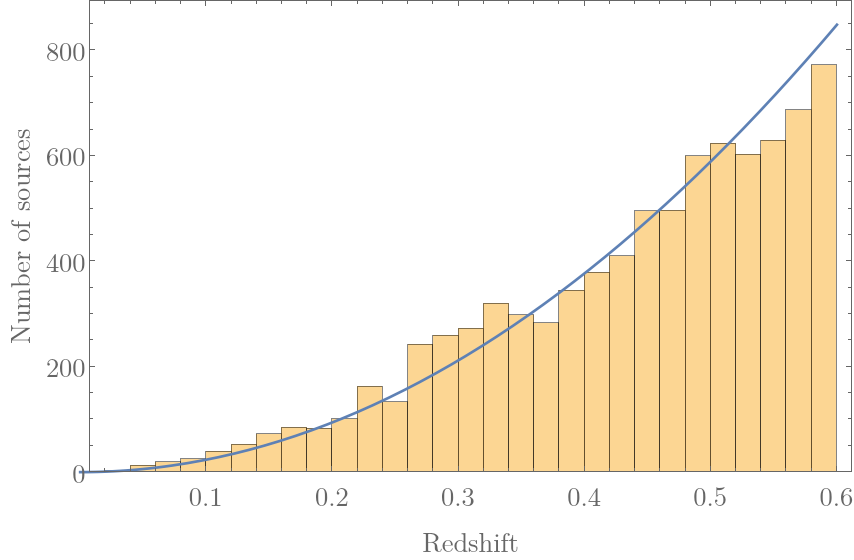}
\includegraphics[width=90mm]{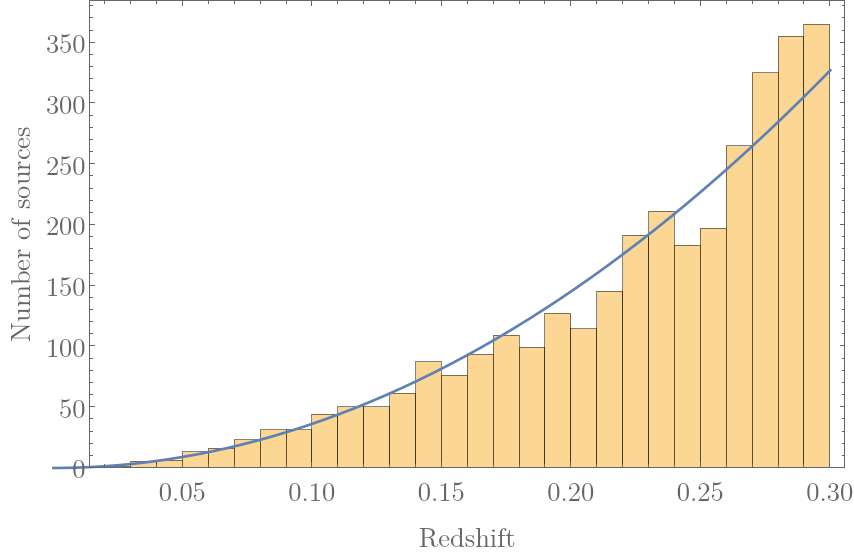}
\includegraphics[width=90mm]{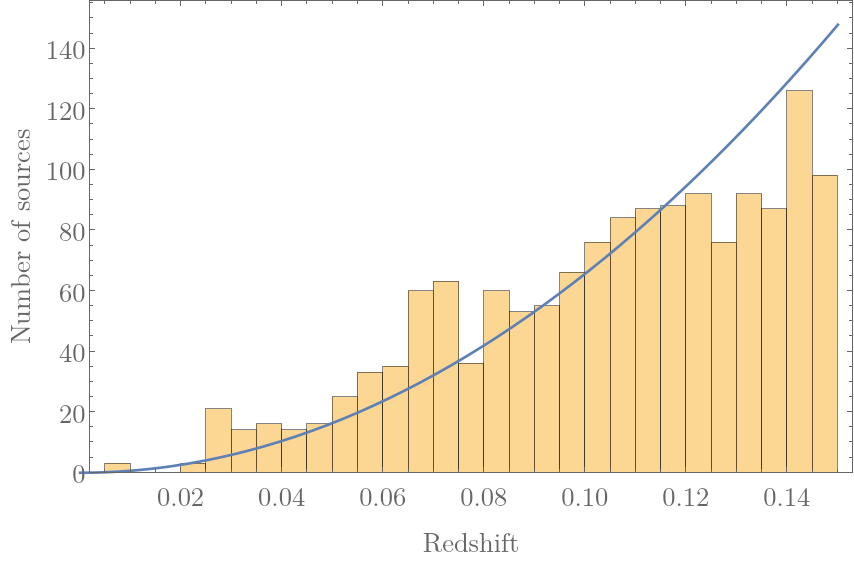}
\end{center}
\caption{Histogram of the redshifts from complete sub-samples of the observations: $\{z < 0.6, L_\nu > 6.63\times10^{23} \, \text{W/Hz}\}$, $\{z < 0.3, L_\nu > 1.35\times10^{23} \text{W/Hz}\}$ and $\{z < 0.15, L_\nu > 1.24\times10^{22} \text{W/Hz}\}$ respectively. We also plot in blue curves the quadratic distribution $P(z)\sim z^2$, first normalized between each redshift interval, and then multiplied by the area of each histogram.}
\label{zquadratic}
\end{figure}

The redshift of the simulated sources is randomized according to the distribution observed in LoTSS. To estimate the redshift distribution from the observations we select RGs unaffected by the flux limit, in order to avoid selection effects. By substituting $ S_\nu = 0.5 \, \text{mJy} $ into expression \eqref{flux}, we can obtain the minimum observed luminosity detectable at each redshift. To determine the intrinsic source luminosity ($ L_\nu $), we assume a fixed intermediate value of $\alpha = 0.75$, since it varies approximately between $0.5$ and $1$ (see Paper I). We find that, in the region $ z < 0.6 $, galaxies with luminosity above $6.63\times10^{23} \, \text{W/Hz}$ at $150~\text{MHz}$ are not affected by the flux limit. In other words, the sub-sample $\{z < 0.6, L_\nu > 6.63\times10^{23} \, \text{W/Hz}\}$ is complete in terms of total flux. The  redshift distribution for that sub-sample roughly fits a quadratic function, as shown in Fig.~\ref{zquadratic}. When applied to other complete sub-samples, such as $\{z < 0.3, L_\nu > 1.35\times10^{23} \text{W/Hz}\}$ or $\{z < 0.15, L_\nu > 1.24\times10^{22} \text{W/Hz}\}$, the procedure yields a similar result (see Fig.~\ref{zquadratic}). 

We impose this redshift distribution to the simulated RLAGN throughout the whole power distribution, thus assuming that the low power sources follow the same number density evolution as the powerful ones, which should be a good approximation at low redshifts, as we do not observe significant changes in the RLF for the observed sample below $z=0.6$ \citep{2019A&A...622A..12H}. We thus select the redshift of the sources from a quadratic probability distribution $P(z)\sim z^2$. 

\subsection{Parameters that define the RGs} \label{param}

The RGs are characterized by a series of parameters (see Paper~I): the jet power, the galactic activity time, observing time/age, the jet viewing angle, the shock dimensions at the starting time of the simulations ($t_0$), the mass-load rates, the cocoon pressure evolution exponents, the adiabatic index, and the electron spectral distribution. All of these parameters are either fixed at a selected value or derived from specific probability distributions, which we describe in detail in this section.

We start the evolution and the injection of particles into the radio-lobe or cocoon at $t_0 = 1000$~yr. The initial ratio between the cocoon's length and radius is defined as
\be
R_{c,0} = 
\begin{cases} 
4 & \text{if }  L_0 \leq 10^{37} \, \mathrm{W}, \\ 
5 & \text{if } L_0 > 10^{37} \, \mathrm{W},
\end{cases}
\ee
$L_0$ being the jet power. We have taken those values following Paper~I, but we don't expect them to play a significant role in the long-term calculated evolution, because this aspect ratio evolves, adapting to the source dynamics. When still compact, elongated sources have higher surface-brightness than wider ones, but at these scales, the main limitation for detection is the flux sensitivity. The particle energy distribution exponent is fixed to $p = 2.14$ as in \citet{1997MNRAS.292..723K}. This exponent determines the amount of non-thermal particles injected into the lobes per unit energy during the history of the source. Therefore, it has an influence on the number of them that will have cooled to the point of having the Lorentz factor that allows them to emit at the given frequency at the observing time (Eqs.~22-28 in Paper~I). The steeper the slope (the larger $p$) is, the less energetic particles injected prior to the observation will be able to contribute to emissivity at a given moment, so the emission becomes more dependent on recently injected particles with small Lorentz factors.
The adiabatic index in the radio lobes is fixed to $\Gamma=5/3$, which accounts for the effects of adiabatic cooling plus the generalized contribution of the mass-loaded cold gas. This is a reasonable assumption for a population dominated by low-energy particles. However, we do not expect significant changes if we use, e.g., $\Gamma=13/9$, the value that corresponds to relativistic electrons and cold protons.
Finally, the viewing angle to the radio source axis, $\theta$, is randomly chosen from a uniform distribution of the quantity $1 - \cos \theta$ between 0 and 1, which results in a uniform distribution of jet orientation in space \citep{2006PhDT........38B}.

\subsubsection{Jet mass-load}

The mass loading of the jet enters in the evolution model of RGs by increasing the thermal energy density in the cocoon. Regarding this,
we defined in Paper~I the evolution of the ratio of thermal to relativistic electron energy densities in the cocoon,
$k(t)=u_{\rm{T}}(t)/u_{\rm{e}}(t)$, as
\be \label{k_paperI}
k(t)= \begin{cases} \displaystyle{c_k\frac{t-t_0}{\sqrt{L_0}}} & \text { if } t \leq t_{3a}, \\ \displaystyle{c_k\frac{t_{3a}-t_0}{\sqrt{L_0}}} & \text { if } t>t_{3a}, \end{cases}
\ee
where $c_k$ is a constant 
and $t_{3a}$ is the time at which the jet's head reaches three times the galactic core radius ($d(t_{3a}) = 3\,a_g$). 

\begin{figure}[htbp]
\begin{center}
\includegraphics[width=90mm]{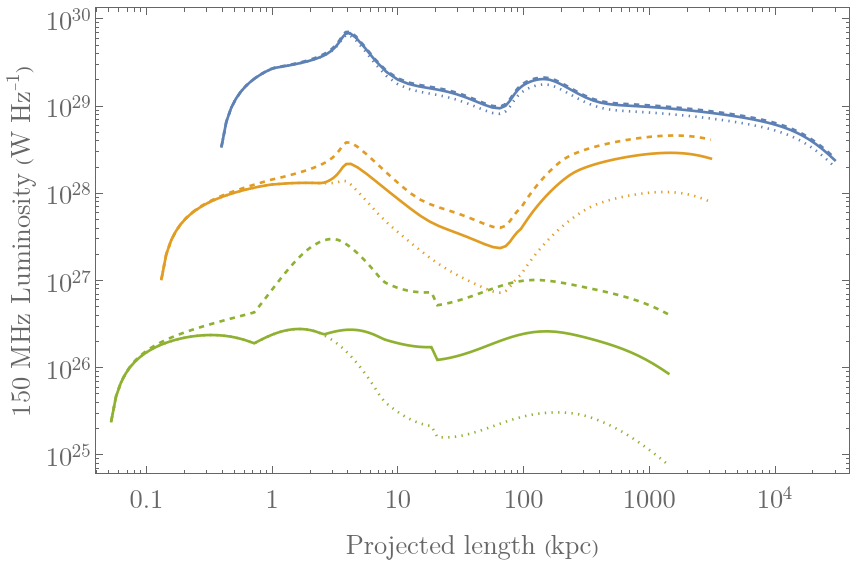}
\end{center}
\caption{Evolution of luminosity vs projected length for different jet powers: $10^{39}$W (blue), $10^{37}$W (orange) and $10^{35}$W (green); and different functions of $k(t)$: Eq. \eqref{k_paperI} with $c_k=10^6W^{1/2} s^{-1}$ (dashed), Eq. \eqref{k_paperI} with $c_k=10^7 W^{1/2} s^{-1}$ (dotted) and Eq. \eqref{k_paperII} (solid). The rest of the parameters are fixed to the following intermediate values: $a_{g} = 1.25$~kpc, $\beta_g = -1.8$, $\rho_{g,0} = 1.67\times 10^{-21}\,{\rm kg/m^{3}}$, $a_c = 45$~kpc, $\beta_c = -1.3$, $\rho_{c,0} = 1.67\times 10^{-23}\,{\rm kg/m^{3}}$, $b_1 = b_5 = -0.85$, $b_2 = -1.85$, $b_4 = -1.475$, $z=0$, and $\theta=90^\circ$.}
\label{kevolucio}
\end{figure}

In this work, we have studied several options for $k(t)$ (see Appendix~\ref{app1}). In particular we have considered the broken linear function defined in Eq.~(\ref{k_paperI}) with two different values of $c_k$, $10^6 \text{\,W}^{1/2}\text{\,s}^{-1}$ (the value used in Paper~I) and $10^7 \text{\,W}^{1/2}\text{\,s}^{-1}$, plus a double-broken linear function
\be \label{k_paperII}
k(t)= \begin{cases} \displaystyle{c_{k,1}\frac{t-t_0}{\sqrt{L_0}}} & \text { if } t \leq t_{a}, \\ 
\displaystyle{c_{k,1}\frac{t_a-t_0}{\sqrt{L_0}}+c_{k,2}\frac{t-t_a}{\sqrt{L_0}}} & \text { if } t_a < t \leq t_{3a}, \\ \displaystyle{c_{k,1}\frac{t_a-t_0}{\sqrt{L_0}}+c_{k,2}\frac{t_{3a}-t_a}{\sqrt{L_0}}} & \text { if } t>t_{3a}, \end{cases}
\ee
that accounts for the change in the mass-load rate when the jet exits the galaxy core. As we will explain later, we obtained the best fits for the double-broken linear function with $c_{k,1}=10^7 \text{\,W}^{1/2}\text{\,s}^{-1}$ and $c_{k,2}=10^6 \text{\,W}^{1/2}\text{\,s}^{-1}$. These are the values and functional form of $k(t)$ that we use to derive the results presented in Sect.~\ref{results}. Figure~\ref{kevolucio} shows the luminosity evolution for radio sources with different powers and different $k(t)$ functions. The rest of the parameters are the same for all the represented cases, and fixed to intermediate values (see the caption).

\subsubsection{Cocoon pressure evolution exponents}

The evolution model is based on the description of the pressure evolution along the five different phases of the cocoon expansion across the ambient medium (Paper~I): the galactic core, a falling galactic density region, the cluster core, a second fall in density, and the transition to the inter-cluster space. In each of these phases the pressure evolves as a power-law of the form $p_c \sim t^{b_i}$, with $i = 1, \ldots, 5$. These exponents are used to describe the pressure evolution during the supersonic phase of the cocoon expansion. Once the shocks disappear and the expansion becomes subsonic, we impose that the pressure evolves with that of the ambient medium and the expansion velocity is the local ambient sound speed. 

When generating a mock population of RLAGN, the values of $b_i$ are uniformly randomized within intervals that depend on the density exponents $\beta_g$ and $\beta_c$, in such a way that we ensure that the structure neither contracts nor accelerates (see Section~4.1 of Paper~I). For the initial and final phases, with constant ambient density, $b_1$ and $b_5$ are chosen within the interval $[-0.9,-0.8]$. In the phases where the density profile decreases with exponents $\beta_g$ (galaxy region) and $\beta_c$ (cluster region), $b_2$ and $b_4$ are taken within intervals $[\beta_i/2-1,\beta_i]$, with $i=g,c$, respectively. In the transition from the galaxy to the cluster core density, we fit the density distribution as a power law and estimate the exponent $\beta_{gc}$ for $\rho \propto l^{-\beta_{gc}}$. For simplicity, and because this region is not, in general, very extended, we fix $b_3$ at the central value of the corresponding interval ($[\beta_{gc}/2-1,\beta_{gc}]$), i.e., $b_3=(3\beta_{gc} - 2)/4$ (see Paper~I for details).

It is important to recall that this refers to the expansion velocity of the front wave. As explained in Sect.~3 of Paper~I, in this regime we consider that 1) the pressure at the particle injection point (the equivalent to the hot-spot pressure in a supersonic radio galaxy) is the ambient pressure at half the distance between the nucleus and the front wave position, and 2) the lobe pressure is the ambient one at the front wave position. We understand that this simplifying assumption may introduce a source of error in both the lobe pressure ($p_c$ in the model) and the injector one ($p_h$). In the former, due to the difference in pressure between the position of the front wave and the probably detached jet lobe in low-power jets, as it can be seen in numerical simulations \citep[e.g.,][submitted]{2014MNRAS.441.1488P,Hervella26}. In the latter, it would be caused by the particle injection being assigned to a given location (as in the case of a hotspot), while it actually takes place continuously along the jet in the dissipation region and probably beyond. However, this approach allows us to get an estimate of the lobe luminosity following the same methodology as used for supersonic sources in the \citet{1997MNRAS.292..723K} approach.

\subsubsection{Source ages at observing time}

The age of each source at the time of observation is randomized following the procedure explained in \cite{1999AJ....117..677B}, which ensures the spatial homogeneity of the distribution. The source age ($t_{age}$) is randomized between $t_0$ ($1000$~yr, see above) and the maximum activity time of the host $ t_{\text{max}} $, following a uniform distribution in the comoving volume. 

The value chosen for $t_{\text{max}}$ significantly affects the distribution of projected lengths in the simulation. For this reason, we have run simulations using different intervals and probability distributions for this parameter. In Appendix~\ref{app1} we show the results obtained for a common, fixed value of $t_{\text{max}}$ (with values 5, 50 and 500~Myr), and for different distributions (uniform in both linear and logarithmic scales or normal, using different time intervals). In Sect.~\ref{results} we show the simulation that gives the best fit to the observed projected linear size distribution: a uniform distribution in logarithmic scale (i.e., using $\log_{10}(t_{\text{max}})$) between 10 and 200~Myr. Interestingly, this distribution is in agreement with the results obtained recently by \citet{2020MNRAS.496.1706S} and \citet{2025MNRAS.537..343Q}, because it is equivalent to $t_{\rm max}^{-1}$ (see Sect.~\ref{sec:disc}).

In the next paragraphs, we explain the way in which $t_{age}$ is chosen between $t_0$ and $t_{\rm max}$ to produce a uniform distribution in comoving volume. 
First, we introduce the cosmological functions that are used. The scale factor $a(t)$ can be obtained by solving numerically the differential equation
\be
\frac{da}{dt}=a H_0 \left(\Omega_M a^{-3}+\Omega_\Delta\right)^{-1/2},
\ee
with the condition $a(0)=0$. The relation between redshift and time is given by $z+1=1/a(t)$. The radial coordinate given as a function of time is
\be
r(t)=\int_t^{t_{now}}\frac{c}{a(t')}dt',
\ee
where $t_{now}$ is the age of the Universe ($a(t_{now})=1$). Finally, the comoving volume cell between two nearby cosmological times is given by (here we assume the scale factor is approximately constant during the source evolution)
\be
\Delta V_C(t_1,t_2)\approx\frac{4\pi}{3}a(t_1)^3\left(r(t_1)^3-r(t_2)^3\right).
\ee

In order to determine the age of a source when observed, we start by choosing a value of the redshift at the birth time of the source, $ z_{\text{birth}} $, which are distributed as explained in Sect.~\ref{ambient}. 
 We then obtain the cosmological time at the birth of the source, $ t_{\text{birth}} $, corresponding to that redshift. The source is located within the comoving volume cell $ \Delta V_C (t_{\text{birth}} + t_0, t_{\text{birth}} + t_{\text{max}})$, and we observe it at cosmological time $ t_{\text{birth}} + t_{\text{age}}$, with $t_{age}\in(t_0,t_{\rm max})$. Therefore, the volume cell between the source birth and the observation time will be a fraction of the complete cell, that is, 
\be\label{frac_vol}
\Delta V_C(t_{\text{birth}} + t_0, t_{\text{birth}} + t_{\text{age}}) = f_V \Delta V_C(t_{\text{birth}} + t_0, t_{\text{birth}} + t_{\text{max}}),
\ee
with $ f_V \in [0,1] $. We randomize $ f_V $ to distribute the sources uniformly in space. Then, we numerically solve Eq.~\eqref{frac_vol} to obtain $ t_{\text{age}} $. This value and its corresponding redshift are the ones that we introduce into the model.

\begin{table*}[th!]
\centering
\caption{Summary of all parameters used in the simulation and their assignment method.}
\begin{tabular}{l l l}
\hline
Parameter & Assignment Method & Value / Interval \\
\hline
\multicolumn{3}{c}{Host Galaxy and Cluster} \\
\hline
Galaxy core radius, $a_g$ & Randomized (uniform) & $[0.5,2]$ kpc \\
Galaxy core density, $\rho_{g,0}$ & Randomized (uniform) & $[0.1,10]$ protons cm$^{-3}$ \\
Galaxy density exponent, $\beta_g$ & Randomized (uniform) & $[-2,\,-1.6]$ \\
Cluster core radius, $a_c$ & Randomized (uniform) & $[36,60]$ kpc \\
Cluster core density, $\rho_{c,0}$ & Fixed (dependent on $\rho_{g,0}$) & $\rho_{c,0} = \rho_{g,0}/100$ \\
Cluster density exponent, $\beta_c$ & Randomized (uniform) & $[-1.5,\,-1.1]$ \\
ICM density, $\rho_{\rm ICM}$ & Fixed (dependent on $z$) & $(3.345\times10^{-28}\,{\rm kg\,m^{-3}})(1+z)^3$ \\
Adiabatic index (ambient medium), $\Gamma$ & Fixed & $5/3$ \\
Ambient sound speed, $c_s$ & Fixed & $6.84\times10^{2}$ km s$^{-1}$ \\
Redshift, $z$ & Randomized (quadratic distribution) & $[0,0.6]$ \\
\hline
\multicolumn{3}{c}{Radio galaxy} \\
\hline
Initial time, $t_0$ & Fixed & $1000$ yr \\
Initial cocoon aspect ratio, $R_{c,0}$ & Fixed (dependent on $L_0$) & $4$ if $L_0\le 10^{37}$ W, $5$ if $L_0> 10^{37}$ W \\
Particle distribution index, $p$ & Fixed & $2.14$ \\
Frequency, $\nu$ & Fixed & $150$ MHz \\
Viewing angle, $\theta$ & Randomized (uniform in $1-\cos\theta$) & $[0,90^\circ]$ \\
Adiabatic index (cocoon), $\Gamma$ & Fixed & $5/3$ \\
Ratio of thermal/magnetic energy density $k(t)$ & Fixed (dependent on $t$ and $L_0$)  & Eq. \eqref{k_paperII} \\
CPEE in the galactic core and ICM, $b_1$, $b_5$\tablefootmark{a} & Randomized (uniform) & $[-0.9,\,-0.8]$ \\
CPEE in the galaxy, $b_2$ & Randomized (uniform) & $[\beta_g/2 - 1,\ \beta_g]$ \\
CPEE in the cluster, $b_4$ & Randomized (uniform) & $[\beta_c/2 - 1,\ \beta_c]$ \\
CPEE in the transition, $b_3$ & Fixed (dependent on $\beta_{gc}$, which is fitted) & $(3\beta_{gc}-2)/4$ \\
Maximum activity time, $t_{\max}$ & Randomized (uniform in log) & $[10,200]$ Myr \\
Source age, $t_{\rm age}$ & Randomized (uniform in volume) & $t_0$ to $t_{\max}$ \\
Jet power, $L_0$ & Randomized (piecewise power law) & $[10^{34.5},10^{38}]$ W \\
\hline
\end{tabular}
\tablefoot{See Sect.~\ref{method_description} for more details. \tablefoottext{a}{CPEE means Cocoon Pressure Evolution Exponent.}}
\label{table_param}
\end{table*}

\subsubsection{Jet power distribution} \label{power_dist}

Finally, prior to starting our simulation, we need to establish the jet power spectral distribution of the simulated sources, $P(L_0)$, in order to drive the randomization within that probability distribution. The whole process is described in Appendix~\ref{app2}.

We tested the possibility of fitting a single power law to the jet-power spectrum, but we were unable to reproduce the luminosity distribution of the LoTSS sample using this approach. Therefore, after testing different possibilities, we decided to fit the jet-power distribution with a piecewise power law by dividing the jet powers between $L_0 = 10^{34.5} \, \text{W} $ and $L_0 = 10^{38} \, \text{W}$ into seven intervals that are equally spaced in the logarithmic scale. For each interval, the distribution is given by $ C_i L_0^{q_i} $, where $ i = 1,2,...7 $, $ q_i $ are the parameters to be determined, and $ C_i $ are constant values chosen to ensure both continuity at the interval limits and normalization of the whole distribution. We applied the Nelder-Mead optimization algorithm to fit the values of $ q_i $. This algorithm is suitable for functions with discontinuities and non-differentiable points. Some approximations were applied with the aim to accelerate the process, as explained in Appendix~\ref{app2}. We repeated the process seven times and with those seven simulations we calculated the mean and the standard deviation of the resulting values. We obtained the following result (see Appendix~\ref{app2} for more details on this process):
\bea
&&q_1 = -3.4 \pm 1.4 \,, \,\,\,\,\,\,\,\,\,\,\,\,q_2 = 2.5 \pm 1.1, \nonumber \\
&&q_3 = -2.04 \pm 0.11 \,,\,\,\,\,\,\,  q_4 = -1.83 \pm 0.17, \nonumber \\
&&q_5 = -0.3 \pm 0.3 \,,\,\,\,\,\,\,\,\,\,\,\,\,  q_6 = -1.0 \pm 0.5, \nonumber \\
&&q_7 = -1.7  \pm 1.4 \,.
\label{paramqi}\eea

\begin{figure}[htbp]
\begin{center}
\includegraphics[width=90mm]{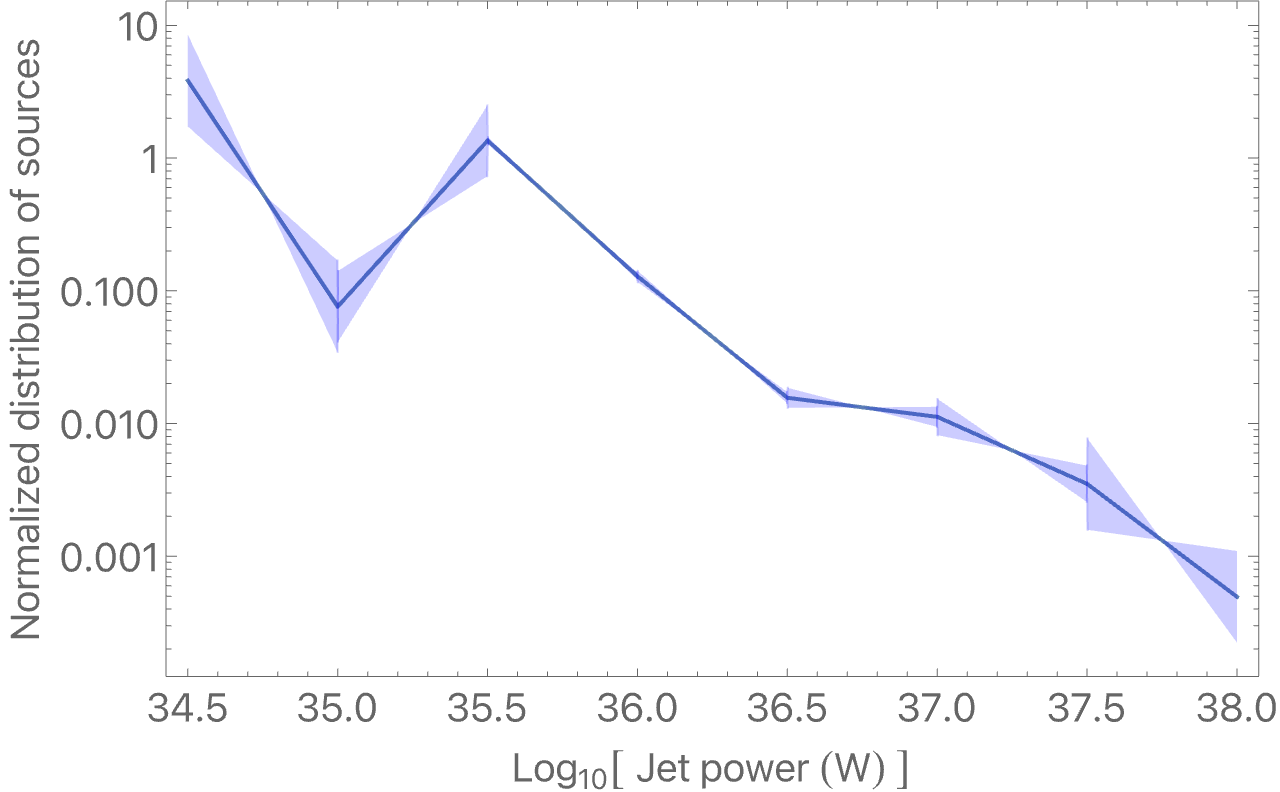}
\end{center}
\caption{Normalized jet power distribution for the set of parameters \eqref{paramqi}. The errors of the slope are represented as a purple shadow.}
\label{distpot}
\end{figure}

Figure~\ref{distpot} shows the power distribution $P(L_0)$ resulting from this set of exponents. It is interesting to note that the uncertainties increase for low and high power sources, where the source numbers are very small. We will discuss this point later in the paper. 

Comparing the exponents of the high-power regime with the result derived by \citet{2025MNRAS.537..343Q} ($P(L_0)\propto L_0^{-1.5}$ for powers between $10^{36.5}$ and $10^{40}$~W, single power-law), we observe a remarkable agreement, despite the different approach followed in both works. The authors point out that they expect a flattening of their distribution at the low power interval they analyze, which we also observe. However, as we will discuss along the paper, both the lower end ($L_0\leq10^{35.5}\,{\rm W}$) and higher end ($L_0\geq10^{37.5}\,{\rm W}$) of our power distribution are affected by different sources of error: absorption within the host galaxy in the case of low power sources, which show typically smaller sizes, and low numbers in the case of the powerful ones (see Appendix~\ref{app2}).

\section{Results}\label{results}

Using all the above parameter distributions and intervals, we generate RGs applying the randomization of the initial parameters. The total number of  sources that we need to simulate in order to obtain the same amount of observable objects as that in the LoTSS sample is also an interesting number in itself, because it gives us information about the possible number of undetected RLAGN there must be for LOFAR to detect the ones it does. We have to consider that compact sources are affected by synchrotron self-absorption and free-free absorption within the host galaxy, which has an impact on observability at low frequencies. Then, by fitting this sample without taking absorption effects into account, we will obtain a lower limit of the number of sources. With this caveat, we proceeded as follows to estimate the total number of RGs: we ran $5$ simulations of $5\times10^5$ RGs each, and calculated the fraction of RGs that would be observable (i.e., those with $S_\nu \geq 0.5 \, \text{mJy}$ and surface brightness above 0.1 $\text{mJy}$) in each case. We obtained the following results in percentage: $\{4.51, 4.48, 4.52, 4.50, 4.47\}$. Taking the mean and the standard deviation of the results, we obtain that the observable fraction is $(4.50 \pm 0.02)\%$. 

As a consequence, to obtain the same number of observable sources as in the LoTSS sample ($17045$), we need a complete sample of (at least) $(3.788 \pm 0.017)\times 10^5$. To construct this sample, we randomly select $17045$ observable sources from the total $2.5\times10^6$ simulated sources obtained by merging the 5 aforementioned sets. We will refer to this set as the observable simulated sample (OSS) from now on. Then we select $3.788\times10^5 - 17045$ non-observable sources, which we will refer to as the non-observable simulated sample (NSS). The union of both sets is called the complete simulated sample (CSS). These are the samples analyzed in the following subsections. 

\subsection{Observable fraction of the simulated sample}

\begin{figure}[htbp]
\begin{center}
\includegraphics[width=90mm]{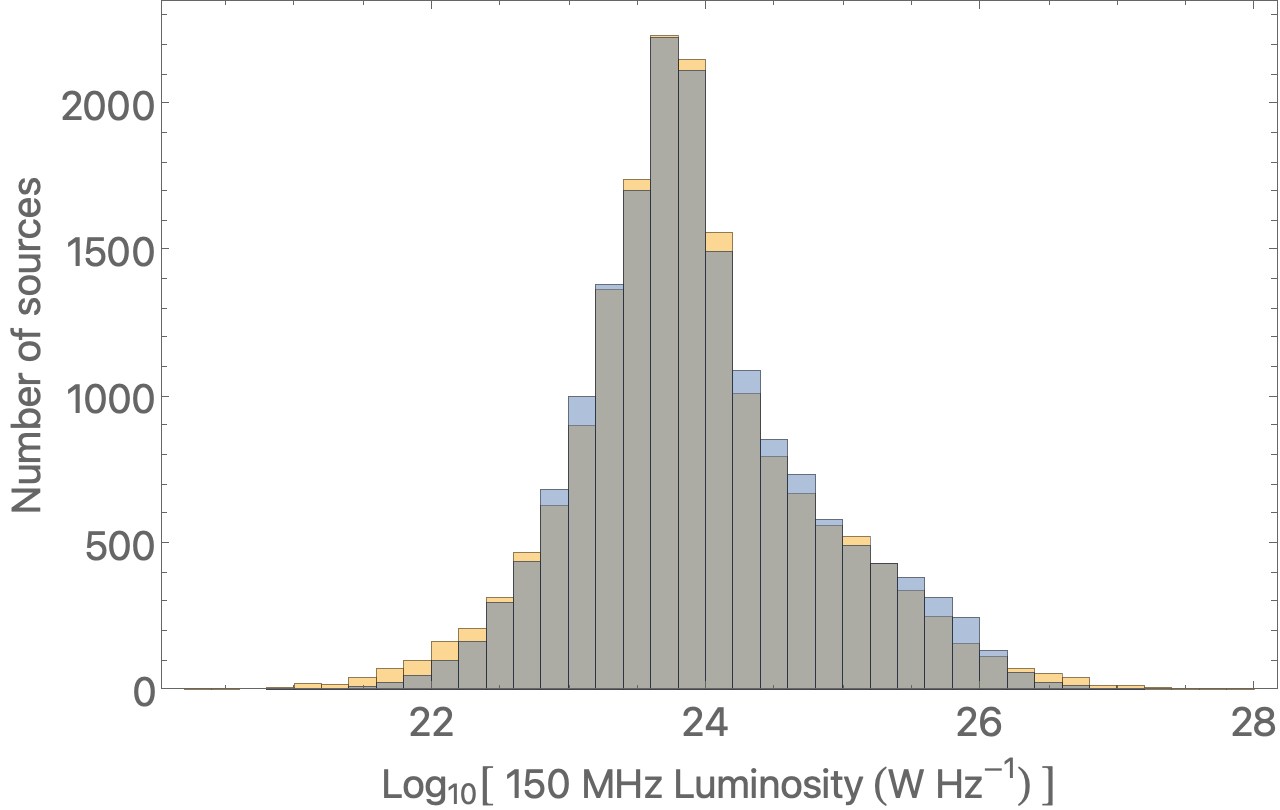}
\end{center}
\caption{Histogram of source luminosities. We show observational sample (LoTSS) in orange and the simulated sample (OSS) in blue.}
\label{histL}
\end{figure}

\begin{figure}[htbp]
\begin{center}
\includegraphics[width=90mm]{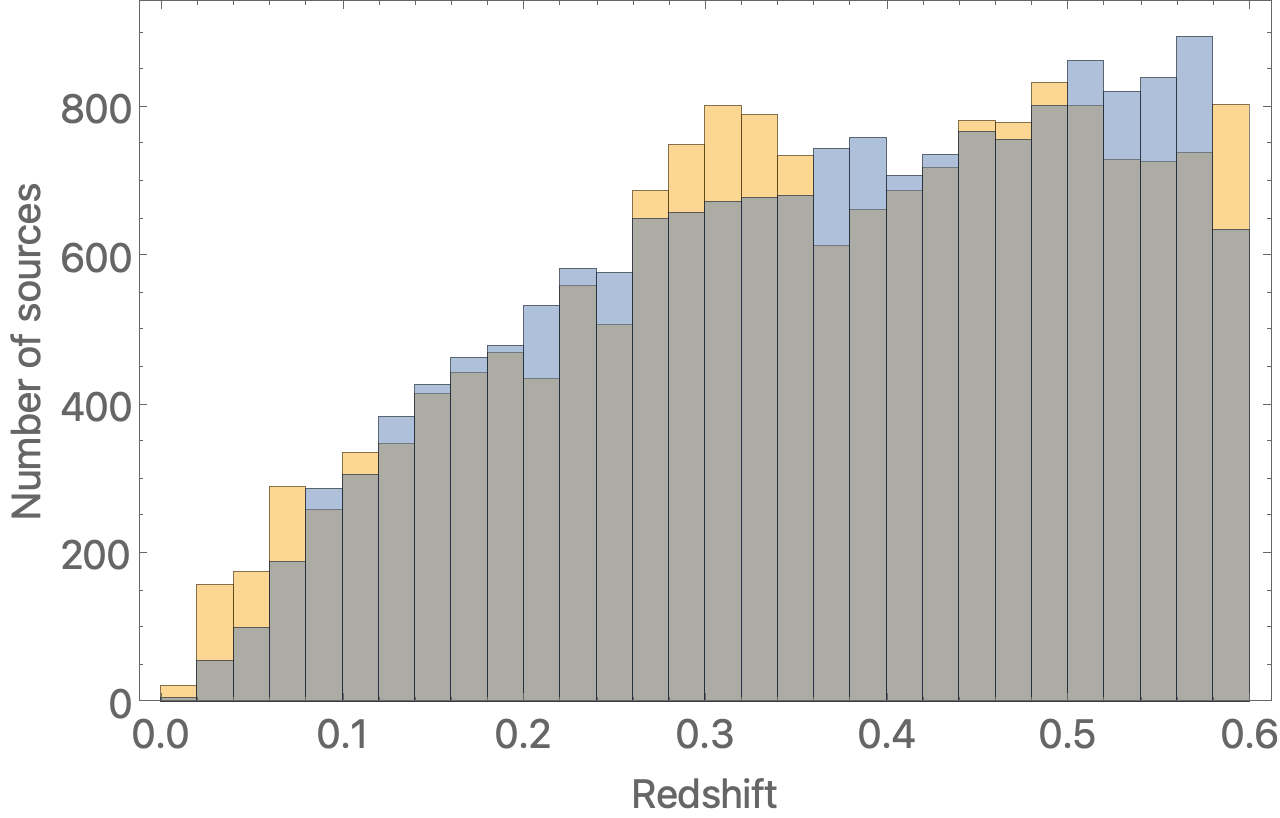}
\end{center}
\caption{Histogram of redshifts. We show observational sample (LoTSS) in orange and the simulated sample (OSS) in blue.}
\label{histz}
\end{figure}

\begin{figure}[htbp]
\begin{center}
\includegraphics[width=90mm]{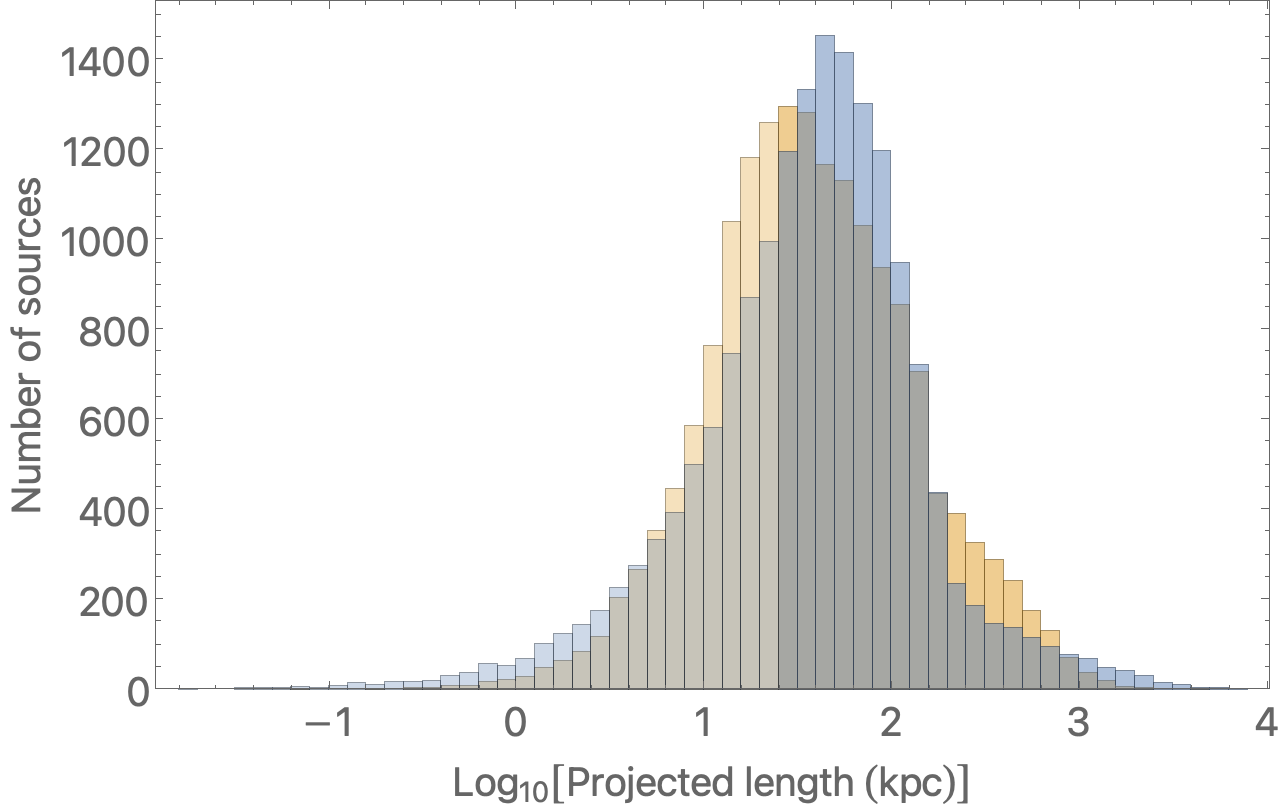}
\end{center}
\caption{Histogram of projected length. We show observational sample (LoTSS) in orange and the simulated sample (OSS) in blue. The light colour indicates the region with sources with sizes $\leq 25~{\rm kpc}$, which can be affected by absorption.}
\label{histD}
\end{figure}

\begin{figure}[htbp]
\begin{center}
\includegraphics[width=90mm]{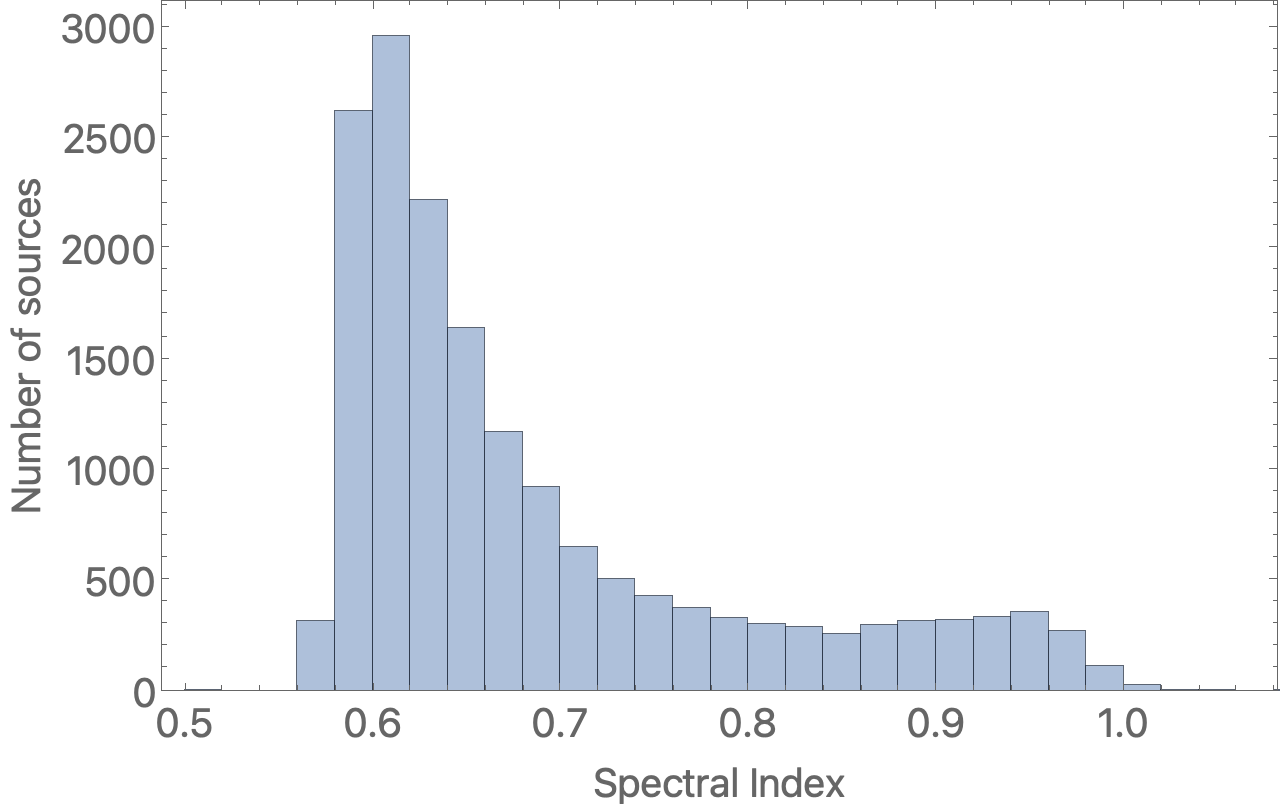}
\end{center}
\caption{Histogram of the radio-lobe spectral indexes of the OSS.}
\label{histalpha}
\end{figure}

\begin{figure}[htbp]
\begin{center}
\includegraphics[width=90mm]{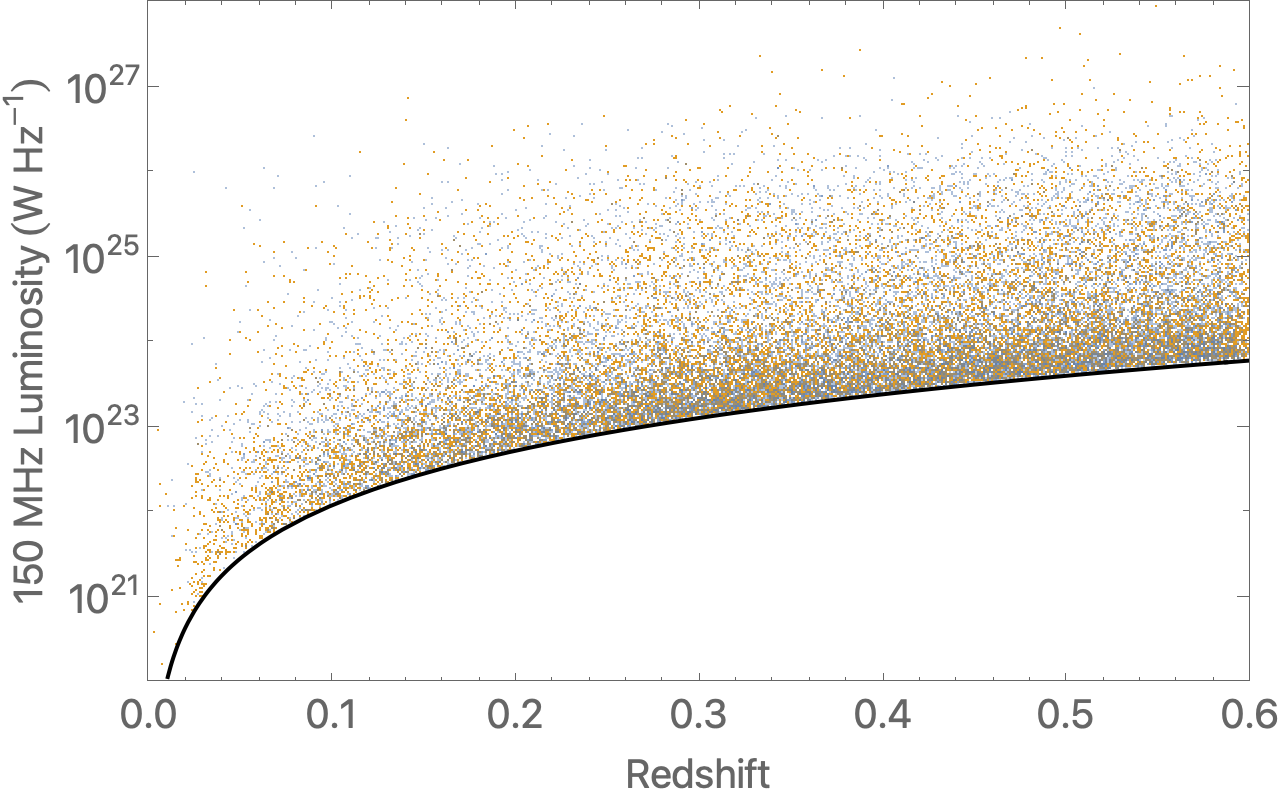}
\end{center}
\caption{Representation of the luminosity versus redshift. In orange the observations (LoTSS) and in blue the simulated sample (OSS). We also show the curve of the minimum observable luminosity due to the flux limitation (black curve). This curve is obtained for the minimum value of the spectral index, which is $(p-1)/2=0.57$.}
\label{Lvsz}
\end{figure}

\begin{figure}[htbp]
\begin{center}
\includegraphics[width=90mm]{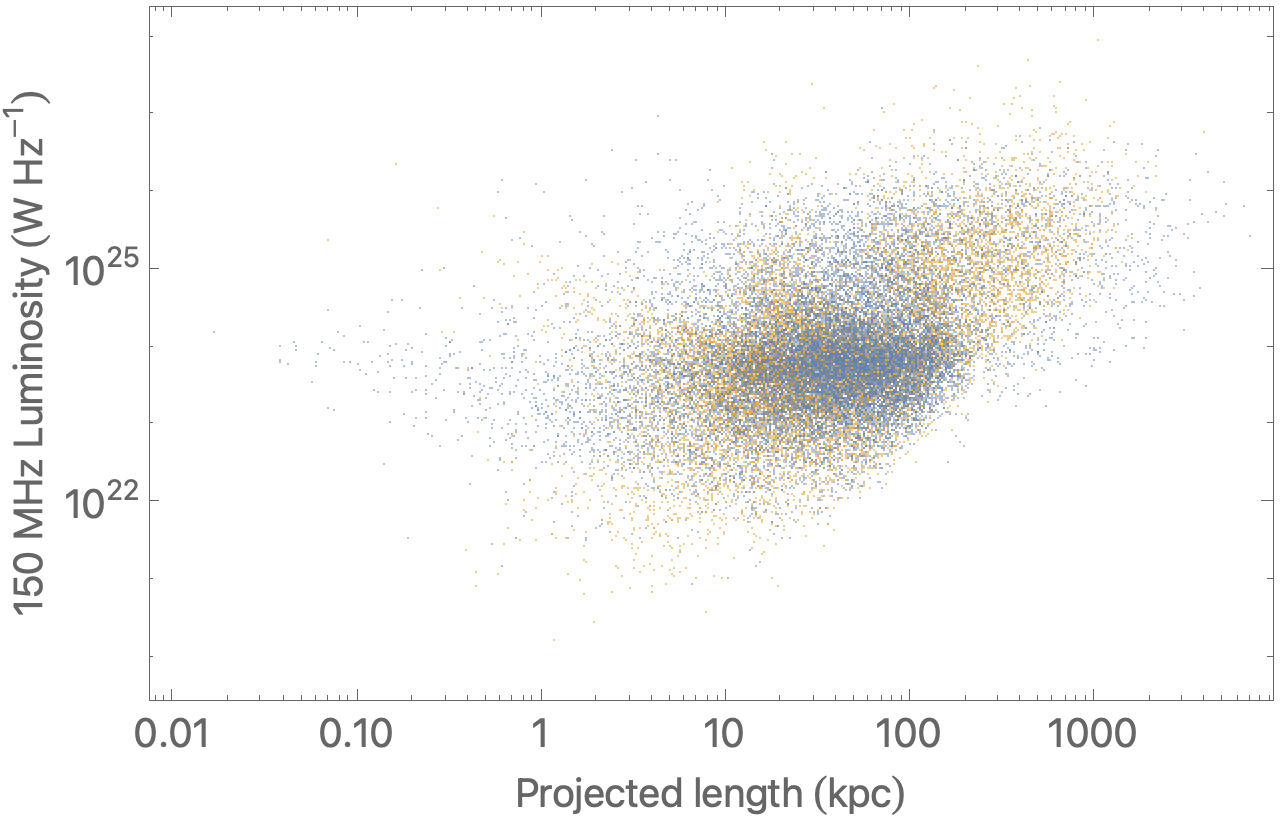}
\end{center}
\caption{Representation of the luminosity versus projected length. We show observational sample (LoTSS) in orange and the simulated sample (OSS) in blue.}
\label{LvsD}
\end{figure}

Figures~\ref{histL}-\ref{LvsD} show the distributions derived for the OSS, by considering the LoTSS sensitivity (a total source flux of $0.5$~mJy and a surface brightness limit of $0.1$~mJy), compared to the distributions given by the RLAGN of the LoTSS sample. The OSS sources are represented in blue in all figures in the paper, whereas the LoTSS sources are always represented in orange. 
 
Figure~\ref{histL} shows the distribution of luminosities for the OSS. The simulation fits almost perfectly with the observations, as a result of a jet power spectrum precisely designed to match the LoTSS luminosity distribution (see Sect.~\ref{power_dist} and Appendix~\ref{app2}). 

In Fig.~\ref{histz} we show the histogram of redshifts. The differences with the observations are small and due to the approximate fit to the observed redshift distribution (see Fig.~\ref{zquadratic}).

Figure~\ref{histD} shows the distributions of the projected lengths for the OSS and the LoTSS samples. In light colour we show the sources with linear sizes $\leq 25$~kpc, which are probably affected by absorption and for which comparison is thus difficult. The model gives a similar distribution to the observed one, which is a proof of the consistency. However, the match is not perfect, with an excess of simulated sources around 100~kpc, and beyond 1~Mpc, on the one hand, and a deficit between 25 and 50~kpc and between 300~kpc and 1~Mpc, on the other.

Moreover, fitting the length distribution has required deep changes (with respect to Paper~I) in the randomization intervals of the parameters and functions that control the model. 
For this reason, we have had to run different sets of simulations, which led us to the conclusion that the mass-load function and the maximum age distribution are the most relevant parameters. In Appendix~\ref{app1} we show the projected length histograms obtained for different distributions of these magnitudes. We discuss those simulations in Sect.~\ref{sec:disc}.

Figure~\ref{histalpha} displays the histogram of the spectral indexes of the OSS sources, Fig.~\ref{Lvsz} shows the luminosity vs redshift for the OSS, along with the observed sources, and Fig.~\ref{LvsD} gives the $P-D$ diagrams, obtained from the luminosity and projected length distributions presented before. The spectral index distribution shows a vast majority of cases with relatively small values, as expected for compact sources, which also dominate the length distribution (see Fig.~\ref{histD}). In Figure~\ref{Lvsz} we see that the simulated population (blue) gives a plausible distribution, similar to that of the real sources (orange), although we observe a deficit of simulated dots at the highest luminosities. This deficit is also observed in Fig.~\ref{LvsD}, associated to extended sources ($\geq 300\,{\rm kpc}$). We tackle these differences in Sect.~\ref{sec:disc}.

\subsection{Complete simulated sample}\label{sec:css}

\begin{figure}[htbp]
\begin{center}
\includegraphics[width=90mm]{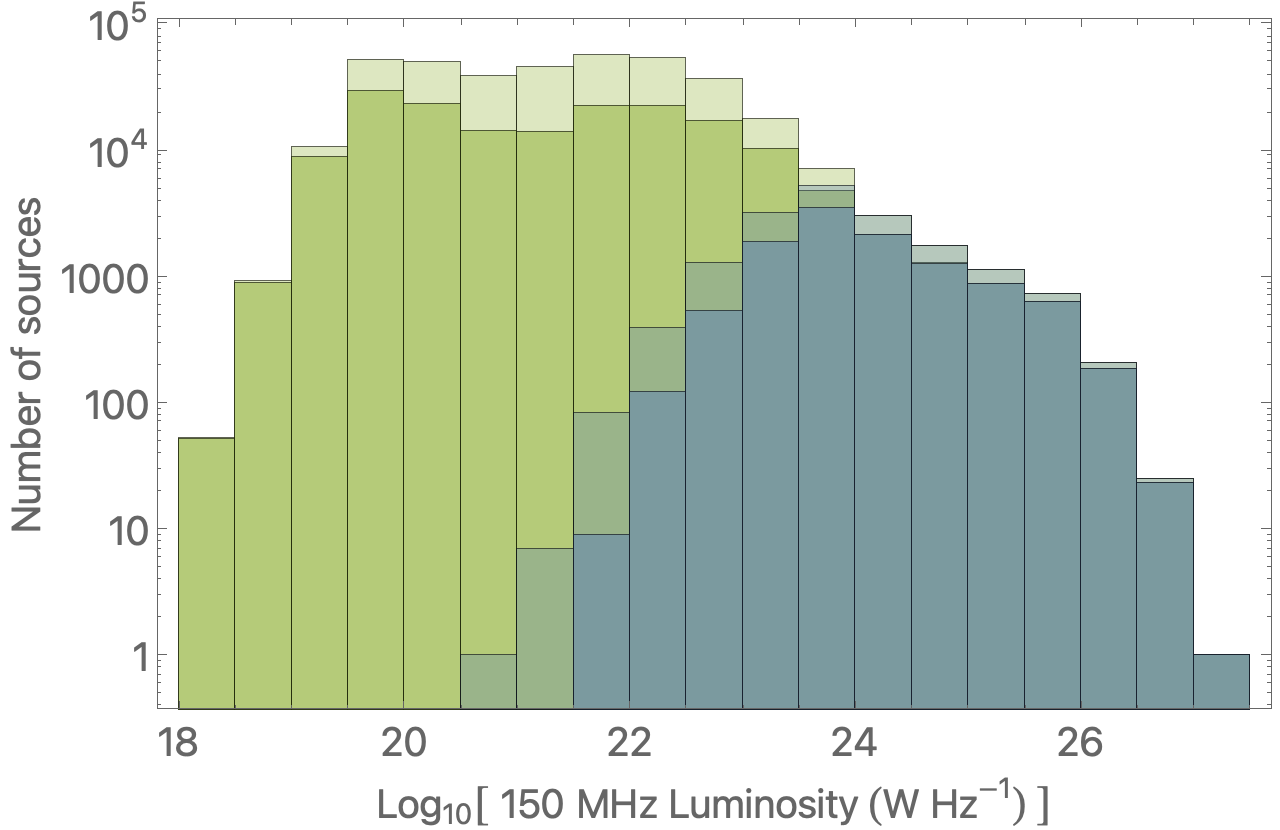}
\end{center}
\caption{Histogram of luminosity of the CSS (green) and the OSS (blue). The light colour columns show the whole sample, whereas the dark colour ones show the same result, excluding all sources with sizes $\leq 25~{\rm kpc}$.}
\label{histLtot}
\end{figure}

\begin{figure}[htbp]
\begin{center}
\includegraphics[width=90mm]{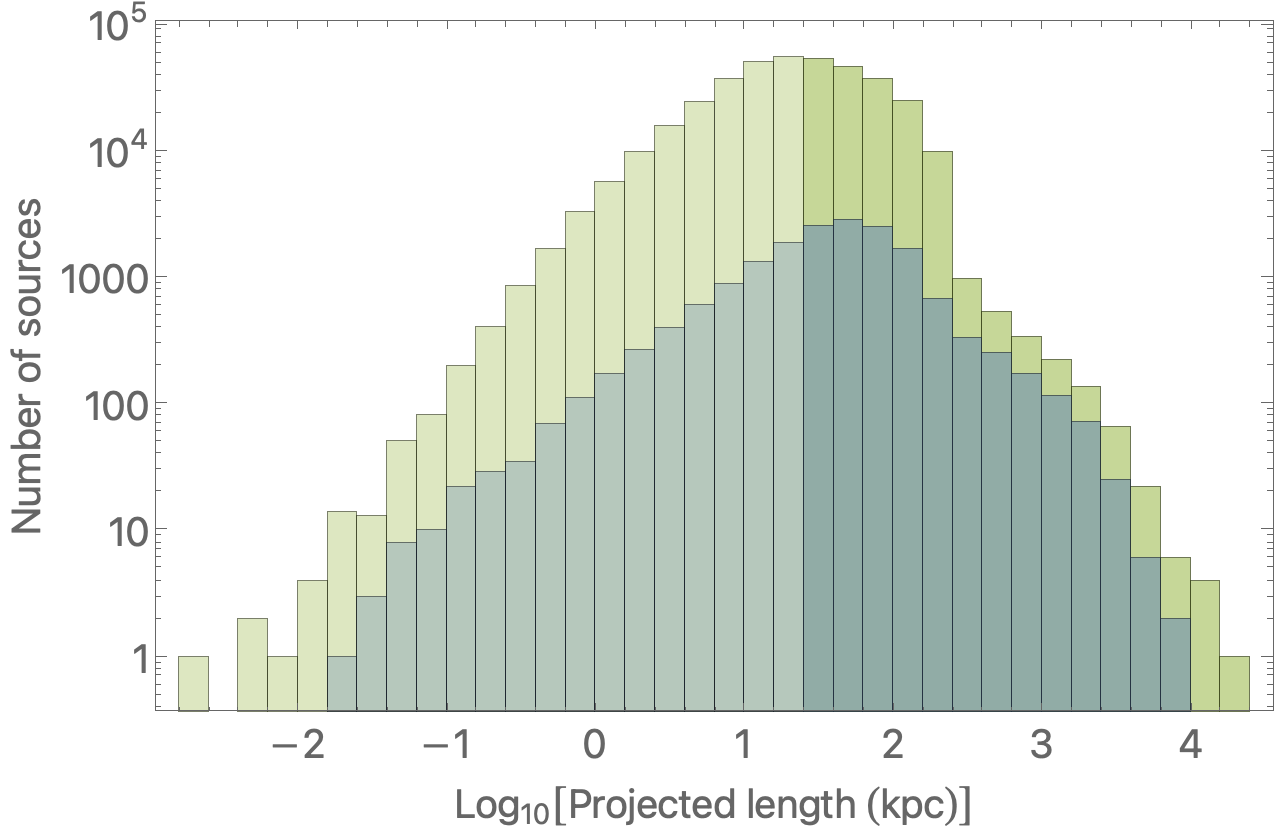}
\end{center}
\caption{Histograms of projected length of the CSS (green) and the OSS (blue). The light colour indicates the region with sources with sizes $\leq 25~{\rm kpc}$, which can be affected by absorption.}
\label{histDtot}
\end{figure}

\begin{figure}[htbp]
\begin{center}
\includegraphics[width=90mm]{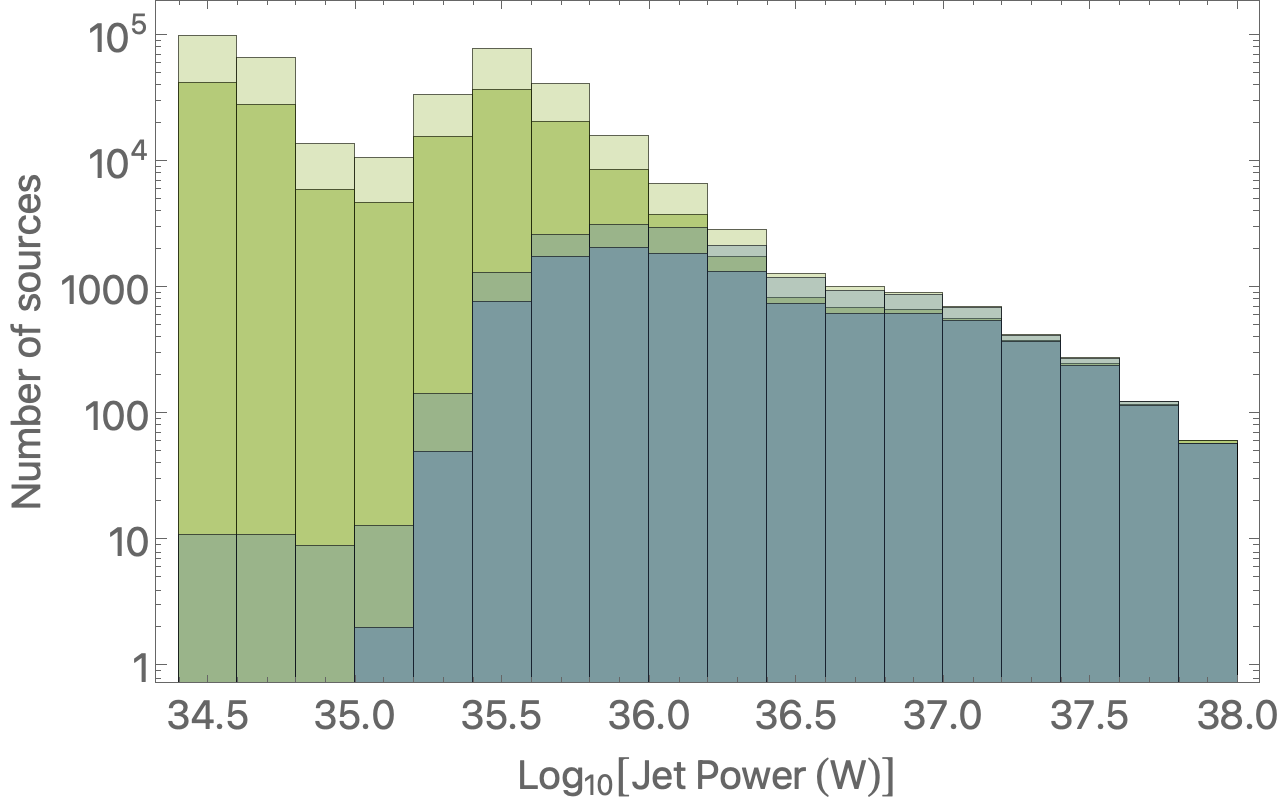}
\end{center}
\caption{Histogram of jet power of the CSS (green) and the OSS (blue). The light colour columns show the whole sample, whereas the dark colour ones show the same result, excluding all sources with sizes $\leq 25~{\rm kpc}$.}
\label{histL0}
\end{figure}

As noted before, we generate a large sample of simulated sources with the jet power distribution required by our model in such a way that we end up producing the same amount of observable RLAGN as given by LoTSS. In this subsection we present the histograms and diagrams of different magnitudes for the CSS (which includes both the OSS and the NSS) compared with the OSS. 

Figures~\ref{histLtot}-\ref{histL0} show the histograms of luminosity, projected length and jet power, with blue columns for the OSS and green ones for the CSS. In Figs.~\ref{histLtot} and \ref{histL0}, the light colour columns show the whole samples, whereas the dark colour ones show only those with sizes above $25~{\rm kpc}$ (i.e., with the tip of the bow shocks beyond the inner $\simeq\,10$~kpc). All histograms are represented in logarithmic scale, since the OSS would be imperceptible using a linear one. Figure~\ref{histLtot} shows that the NSS dominate at low luminosities, as expected. We see that the compact component of the sample is concentrated towards low luminosities and that it barely affects the result for $L_{\rm 150~MHz}>10^{23}\,{\rm W\,Hz^{-1}}$. Figure~\ref{histDtot} shows that a number of extended sources evolve to sizes in which their luminosity makes them non-detectable, as indicated by the green columns being larger than the blue ones. We also observe a steep drop (of about an order of magnitude) in the number of sources at projected distances $D \approx 250$ kpc, caused by the maximum imposed in the source ages ($t_{\rm max}\leq 200$ Myr) and the universal value of the ambient medium sound speed, as discussed below. Finally, Fig.~\ref{histL0} indicates that the NSS population is concentrated at powers $L_0 \leq 10^{37}~{\rm W}$. It also shows that the comparison between the simulated and observed samples can be strongly affected by absorption mainly at the lower end of our power distribution, i.e., for $L_0 \lesssim 10^{36}\,{\rm W}$.

\begin{figure}[htbp]
\begin{center}
\includegraphics[width=90mm]{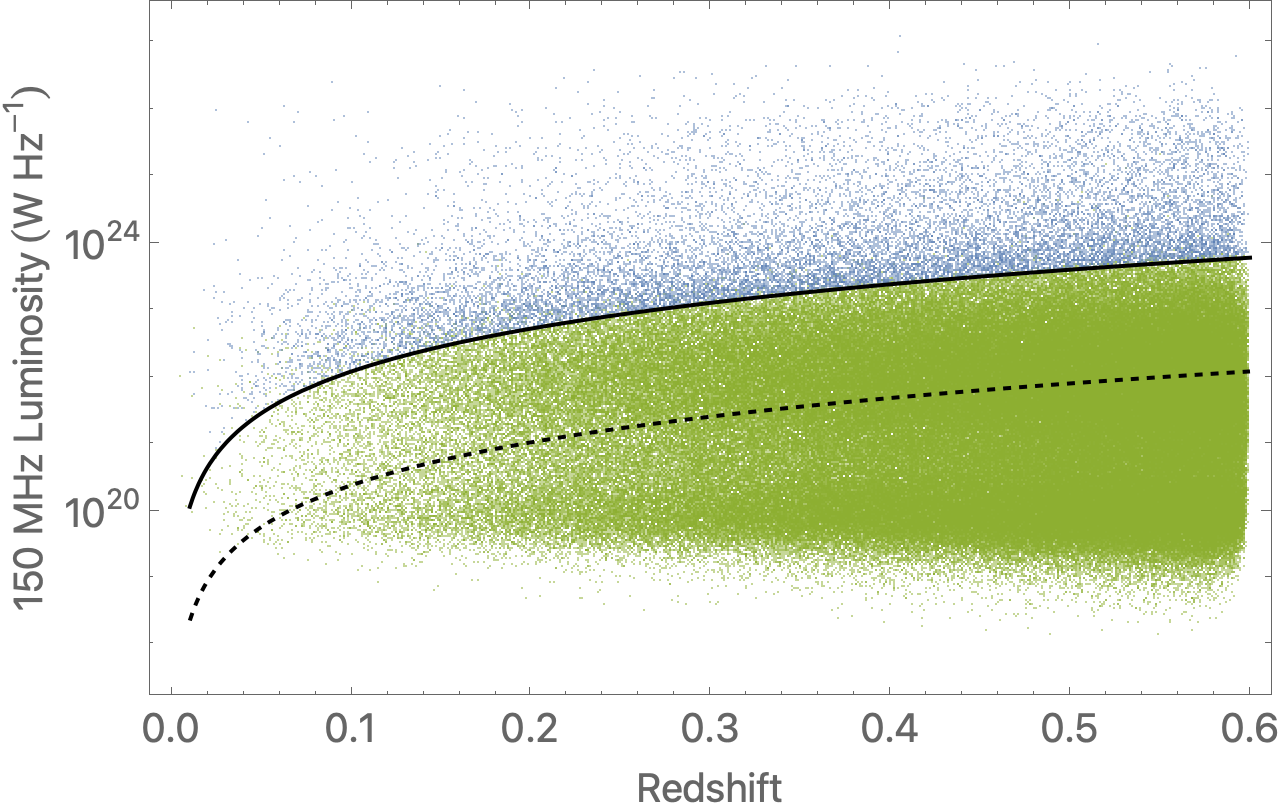}
\end{center}
\caption{Representation of luminosity versus redshift for the NSS (green) and the OSS (blue). The black curve corresponds to the LoTSS sensitivity threshold $S_\nu=0.5$mJy, and the dashed black curve to a threshold of $0.01$mJy, both at the minimum value of the spectral index ($\alpha=(p-1)/2=0.57$).}
\label{Lumvsztot}
\end{figure}

\begin{figure}[htbp]
\begin{center}
\includegraphics[width=90mm]{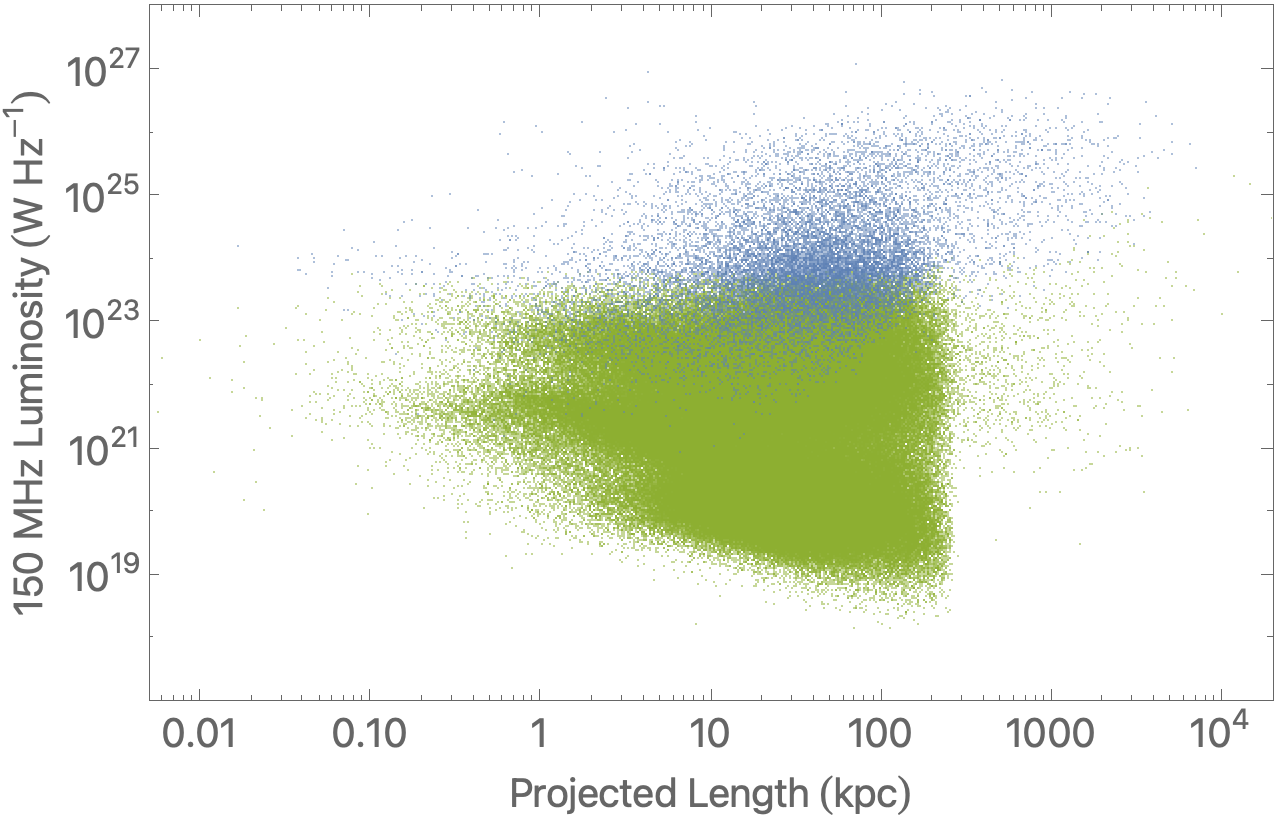}
\end{center}
\caption{Representation of luminosity versus projected length for the NSS (green) and the OSS (blue).}
\label{LumvsDtot}
\end{figure}

\begin{figure}[htbp]
\begin{center}
\includegraphics[width=90mm]{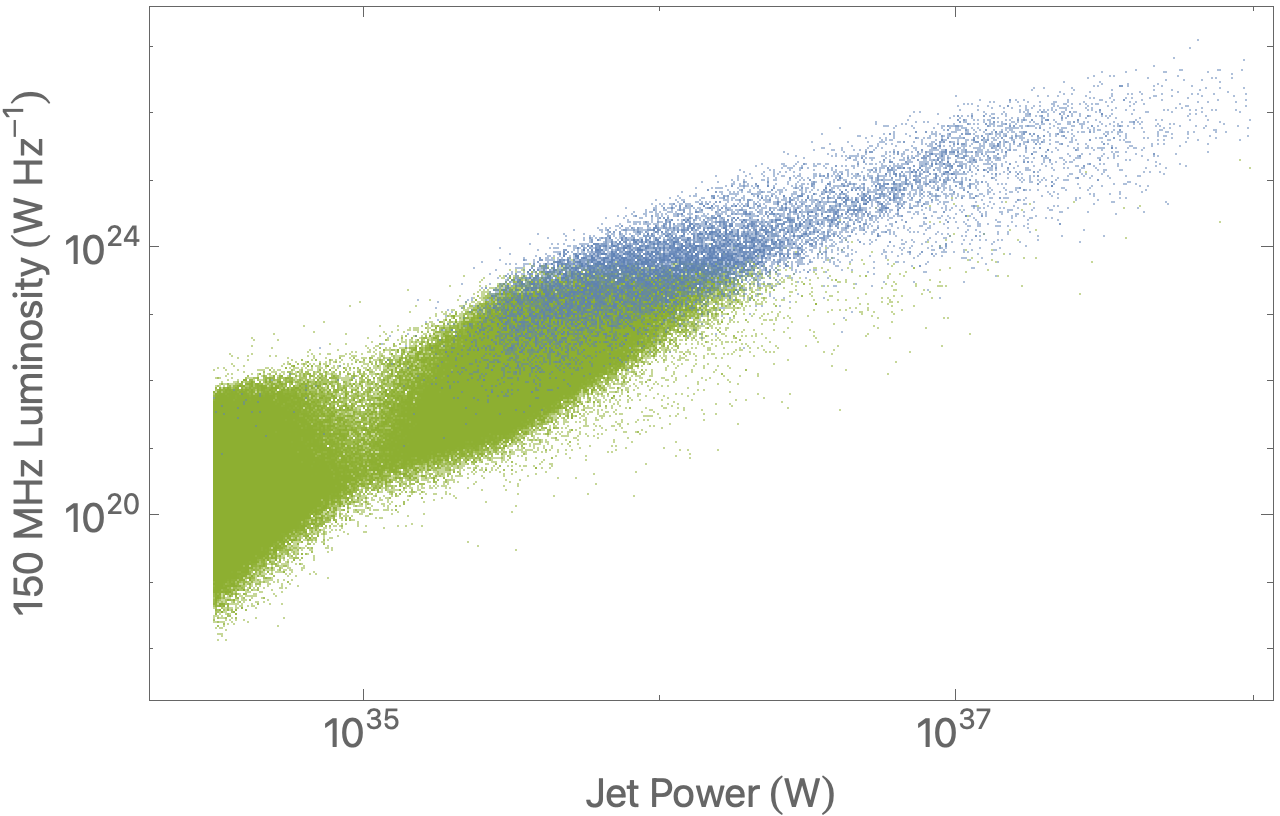}
\end{center}
\caption{Representation of luminosity versus jet power for the NSS (green) and the OSS (blue).}
\label{LvsL0tot}
\end{figure}

\begin{figure}[htbp]
\begin{center}
\includegraphics[width=90mm]{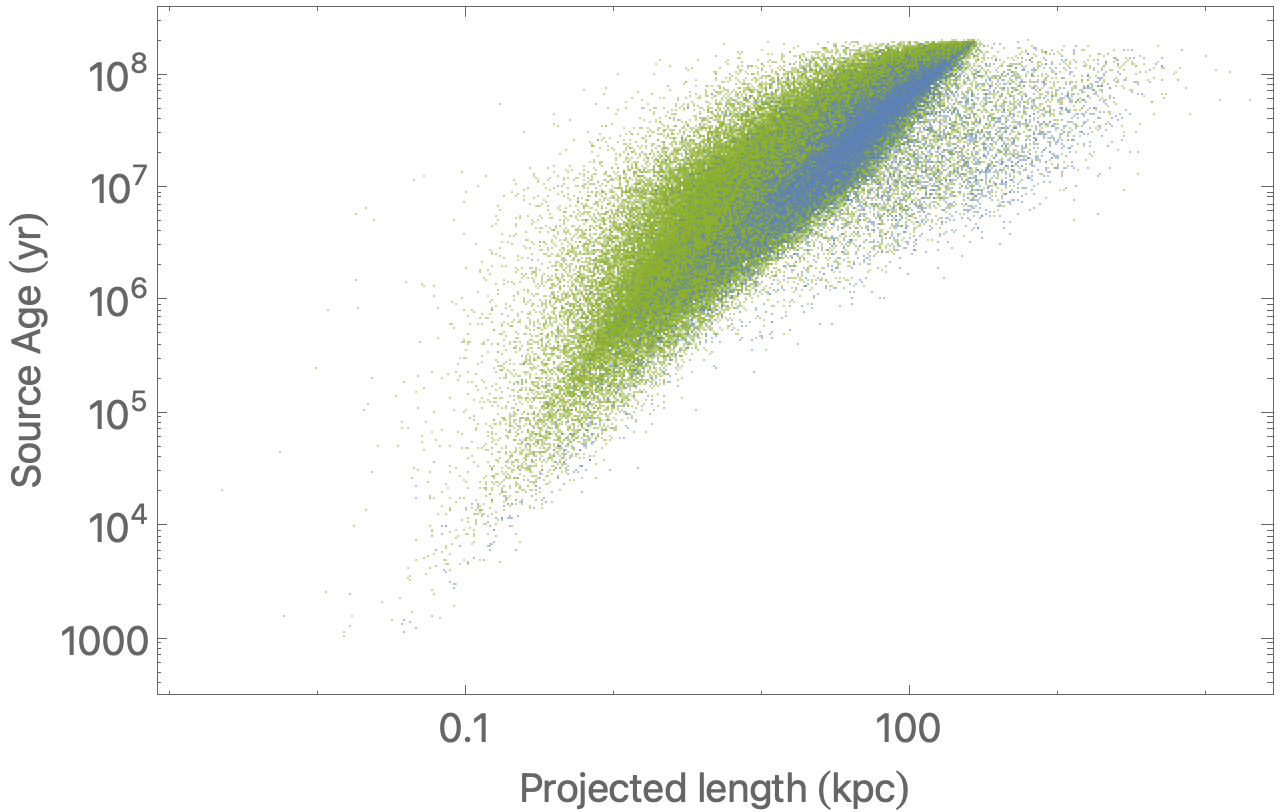}
\end{center}
\caption{Representation of ages of the sources versus projected length for the NSS (green) and the OSS (blue).}
\label{AgevsD}
\end{figure}

Figures~\ref{Lumvsztot}-\ref{AgevsD} show combined plots of different magnitudes for the OSS (blue points) and the NSS (green points). The first plot (Fig.~\ref{Lumvsztot}) shows the luminosity vs redshift distributions above and below the sensitivity threshold and would allow us to make detectability predictions for future arrays, with improved sensitivity. For instance, using this plot we estimate that, within the same solid angle, SKA could detect at least an order of magnitude more sources than LOFAR. Assuming the SKA's sensitivity threshold of $\sim 0.01,\rm{mJy}$ (indicated in the plot by the dashed line), our model predicts at least $\simeq 1.5\times10^5$ detectable sources (this number is obtained without considering the surface brightness limit). However, let us point out that this estimate should be taken as a lower limit and also with certain caution because our estimate can be strongly affected by the caveats of our model (e.g., absorption or an accurate modeling of mass-load) at the low power range.

The $P-D$ diagram in Fig.~\ref{LumvsDtot} shows that the OSS appears shifted towards larger projected lengths with respect to the NSS, because they have, on average, higher jet powers and therefore propagate faster. Regarding the NSS, the sharp break in the source number shown in Fig.~\ref{histDtot} at projected lengths $D \approx 250$ kpc is also seen here. This drop is related to the maximum activity time and propagation velocity of low power sources, as discussed below. We also see a hidden population of giant RGs in Fig.~\ref{LumvsDtot}, probably arising when relatively low power jets propagate through dilute ambient media and can reach large sizes but do not reach the flux or surface brightness thresholds. Finally, the plot reveals a lack of small sources ($D< 2$ kpc) with luminosities between $10^{22}$ and $10^{23}\,{\rm W\,Hz^{-1}}$, probably associated with the drop in the number of sources with low jet powers ($L_0 \approx 10^{35}$ W) in our power distribution (see Fig.~\ref{distpot}).

Figure~\ref{LvsL0tot} shows the luminosity distribution as a function of jet power. Despite the almost log-log linear relationship between the two magnitudes, the width of this distribution reveals how difficult it is to establish a direct correlation between jet power and radio luminosity \citep[an attempt that has been made in different papers, e.g.,][]{2008ApJ...686..859B,2010ApJ...720.1066C,2012MNRAS.423.2498D,2013ApJ...769..129S}, even for a relatively simple, semi-analytical approach. A correlation can perhaps be established for a range of jet powers/luminosities, less affected by, e.g., mass-load, but it breaks below $10^{36}$ and above $10^{37}$~W. We observe that, for a given jet power, the resulting luminosity can vary within up to two orders of magnitude, depending on the environmental conditions and the source age. The scatter could be reduced by including the dependence of luminosity on source size and limiting the study to bright, powerful sources \citep[as done in][]{2013ApJ...769..129S}.

The source distribution in the power-luminosity plane shows a drop in source count at $10^{35}$~W caused by the jet power spectrum (Fig.~\ref{distpot}). This region is, moreover, probably subject to strong absorption effects, because we have seen that a large number of low power sources can have small linear sizes (see Figs.~\ref{histDtot}-\ref{histL0}). This relation between the drop in the jet power spectrum and the source distribution in this plane reveals the sensitivity of our results to changes in the former, mainly at the lower edge, where the number of expected sources increases dramatically. Owing to this increase in the number of sources, a proper evaluation of AGN occurrence and relevance must rely on a proper determination of the power spectrum at low powers, which requires a correct characterization, in particular, of galactic extinction and mass-load. This effort can be certainly pushed by the availability of more sensitive samples.

Finally, Fig.~\ref{AgevsD} shows the source age versus projected length. For the OSS, we see a main set of sources that follows a correlation region throughout two orders of magnitude, but also a large number of outliers. We can interpret these outliers in terms of jet powers: less common, powerful jets require less time to propagate to larger scales. We also see that giant RGs may not require very long ages to reach Mpc scales ($10^7$ to $10^8$~yr), implying mean propagation speeds up to 0.3$\,c$. This can be a robust result if we take into account that those giants show well collimated morphologies, making them more easily reproducible by semi-analytical models \citep{2024Natur.633..537O}. On the extremes, the plot shows, on the one hand, a large number of frustrated sources among the NSS population, with ages $\sim10^7$~yr and sizes of $\sim10\,{\rm kpc}$, and, on the other, a size distribution that extends to Mpc scales. 

Additionally, this plot also shows the large drop in the number of sources with projected lengths beyond $\sim 250$\,kpc, at the maximum source age of $2 \times 10^8$ years, already seen in Figs.~\ref{histDtot} and \ref{LumvsDtot}. The reason for the drop is precisely related to the maximum activity time, and it is produced by the maximum size of the forward perturbations that propagate at the ambient sound speed ($c_s=2.28\times10^{-3}c$) for a large portion of their lives. To test this, we have reproduced these plots (not shown) for different distributions of $t_{\rm max}$: 1) a uniform distribution between 5 and 500~Myr, 2) a uniform logarithmic distribution within the same interval, and 3) a normal distribution centered at 100~Myr and a standard deviation of 50~Myr. We see that all distributions generate such a cut in the NSS. In cases 1 and 2 the cut appears at the same position (600-700~kpc). For case 3, the cut appears at 200~kpc and it is slightly more diffuse. This occurs basically for low power (unobservable) sources because they reach pressure equilibrium with the ambient medium very early in their evolution. Using a range of temperatures for different galactic environments would certainly smoothen the discontinuity; we leave this for future work.

\subsection{Luminosity functions}

In this section, we use the OSS and CSS samples defined before to construct their radio luminosity functions and their jet power function \citep[or jet kinetic luminosity function, as defined in][]{2019A&A...622A..12H}. For this purpose, we define the luminosity function as:

\begin{equation}
\phi(L,z) = \frac{N(L,z)}{V_{\rm C}(z)\,\Delta\log_{10}L},
\end{equation}
where $N(L,z)$ is the number of sources in a given redshift and luminosity bin ($\Delta\log_{10}L$), and $V_{\rm C}(z)$ is the comoving volume of the corresponding redshift shell. For all figures, we split the luminosity functions into six redshift bins, from $z=0.01$ to $z=0.6$.

\begin{figure}[htbp]
\begin{center}
\includegraphics[width=90mm]{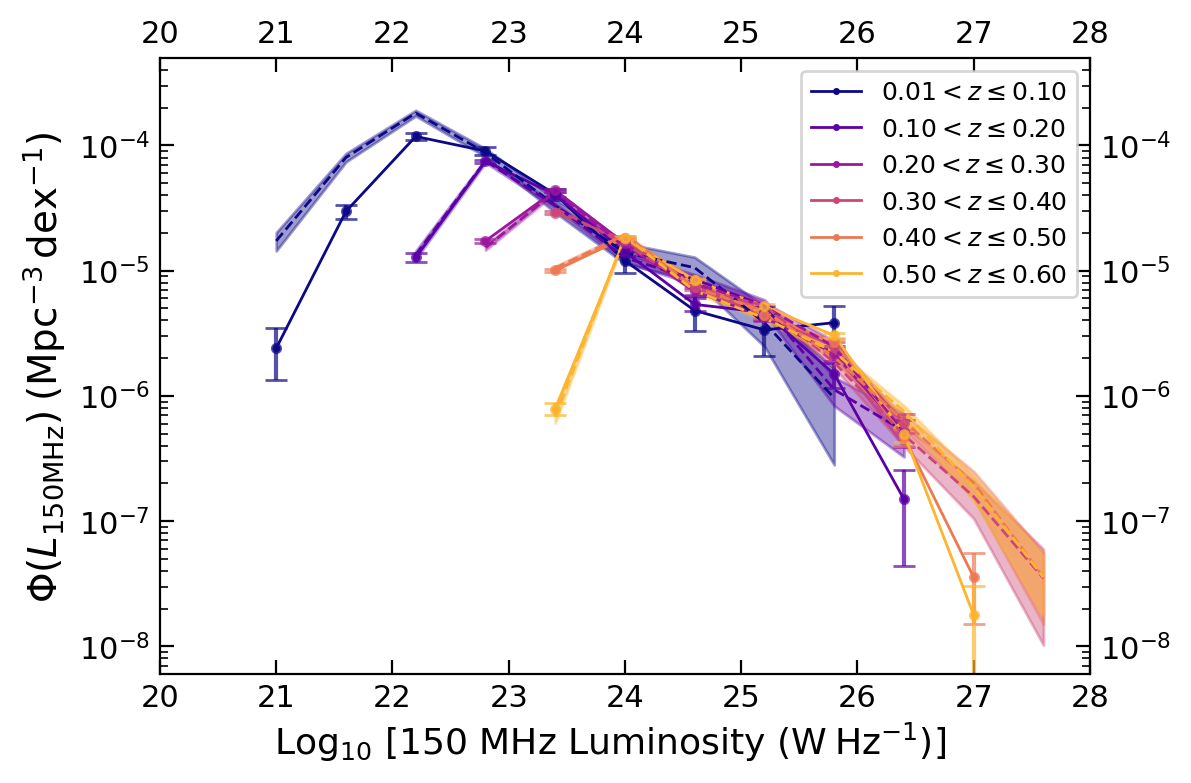}
\end{center}
\caption{Radio luminosity function of the OSS (solid lines and points with error bars) and LoTSS (dashed lines, with errors represented as shaded areas) for different bins of redshift. The errors correspond to the Poissonian uncertainties.}
\label{lum_sim_obs}
\end{figure}

\begin{figure}[htbp]
\begin{center}
\includegraphics[width=90mm]{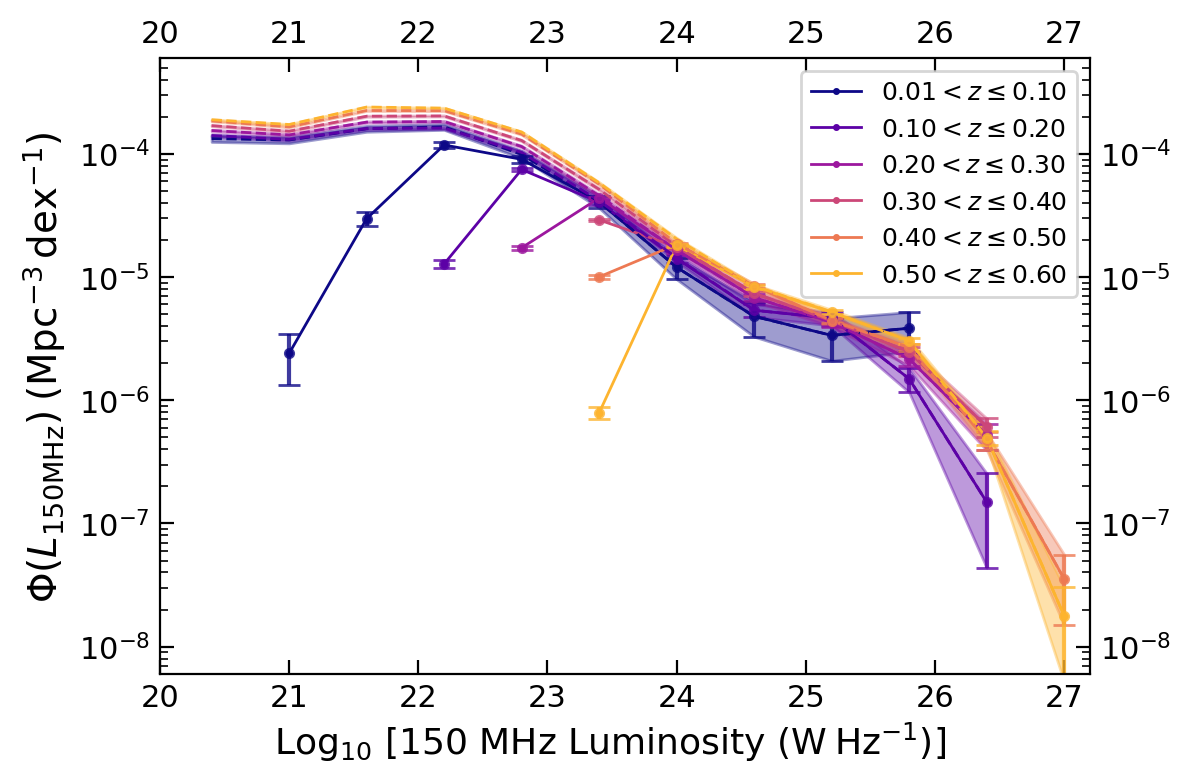}
\end{center}
\caption{Radio luminosity function of the OSS (solid lines and points with error bars) and CSS (dashed lines, with errors represented as shaded areas) for different bins of redshift. The errors correspond to the Poissonian uncertainties.}
\label{lum_sim}
\end{figure}

Figures~\ref{lum_sim_obs} and \ref{lum_sim} show the RLF of the OSS, compared to those of the LoTSS RLAGN sample, and the CSS, respectively. We see from Fig.~\ref{lum_sim_obs} that, overall,  our model reproduces the observed distributions down to the limiting luminosity at each redshift. The distributions deviate more from each other in the regions where the number of sources in both the observations and our model is much lower, and therefore the statistical error is larger. This includes the high and low luminosity ends and the low redshift intervals. The effect of applying a given flux limit ($S_\nu>0.5\,\rm{mJy}$) to the complete sample is clearly visible in Fig.~\ref{lum_sim}, in the large difference between the distributions at the smallest radio luminosities at each redshift bin. This figure shows the potential of this model to make predictions for future surveys with deeper flux limits with respect to LOFAR. 

\begin{figure*}[htbp]
\begin{center}
\includegraphics[width=90mm]{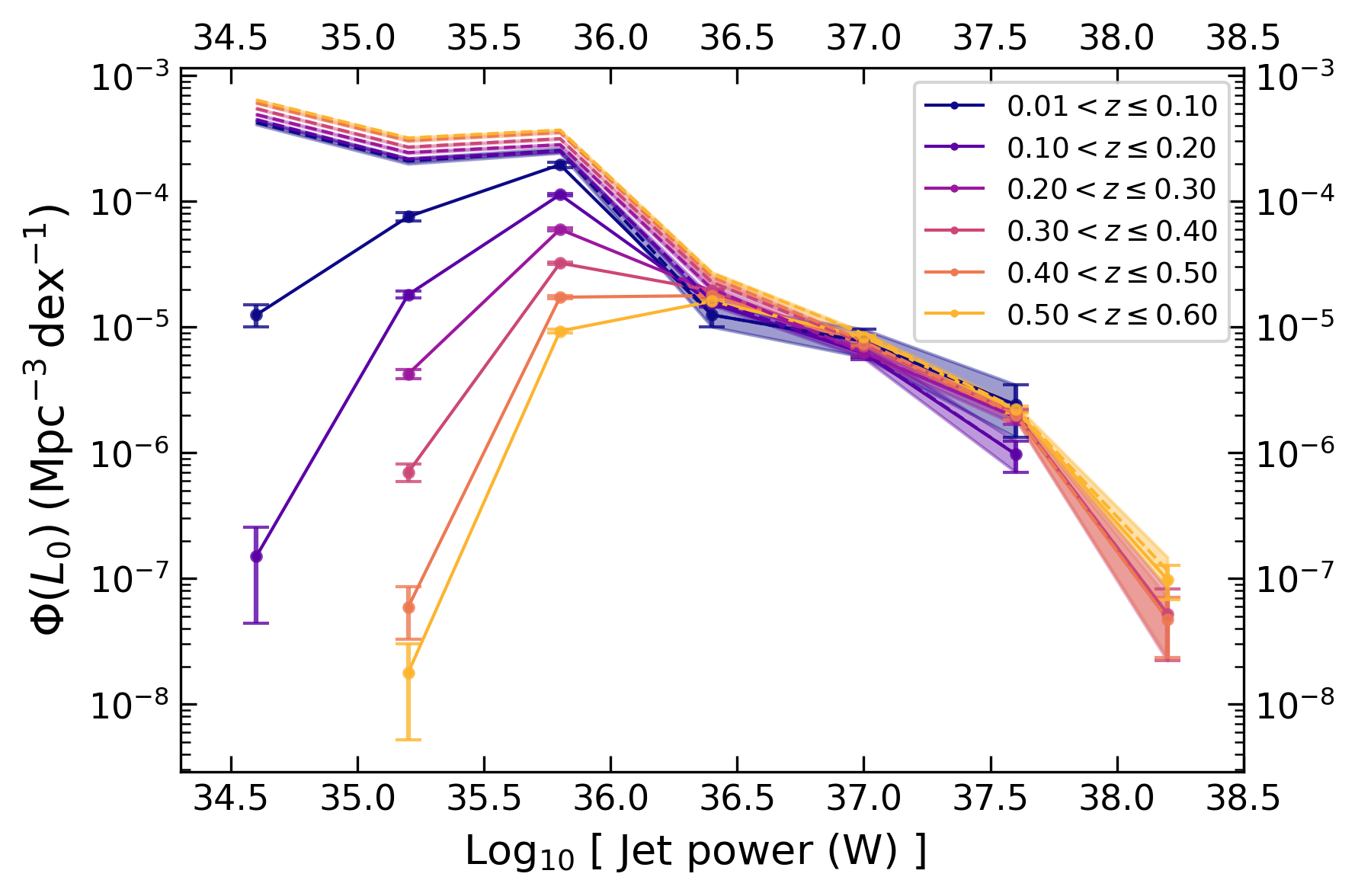}
\includegraphics[width=90mm]{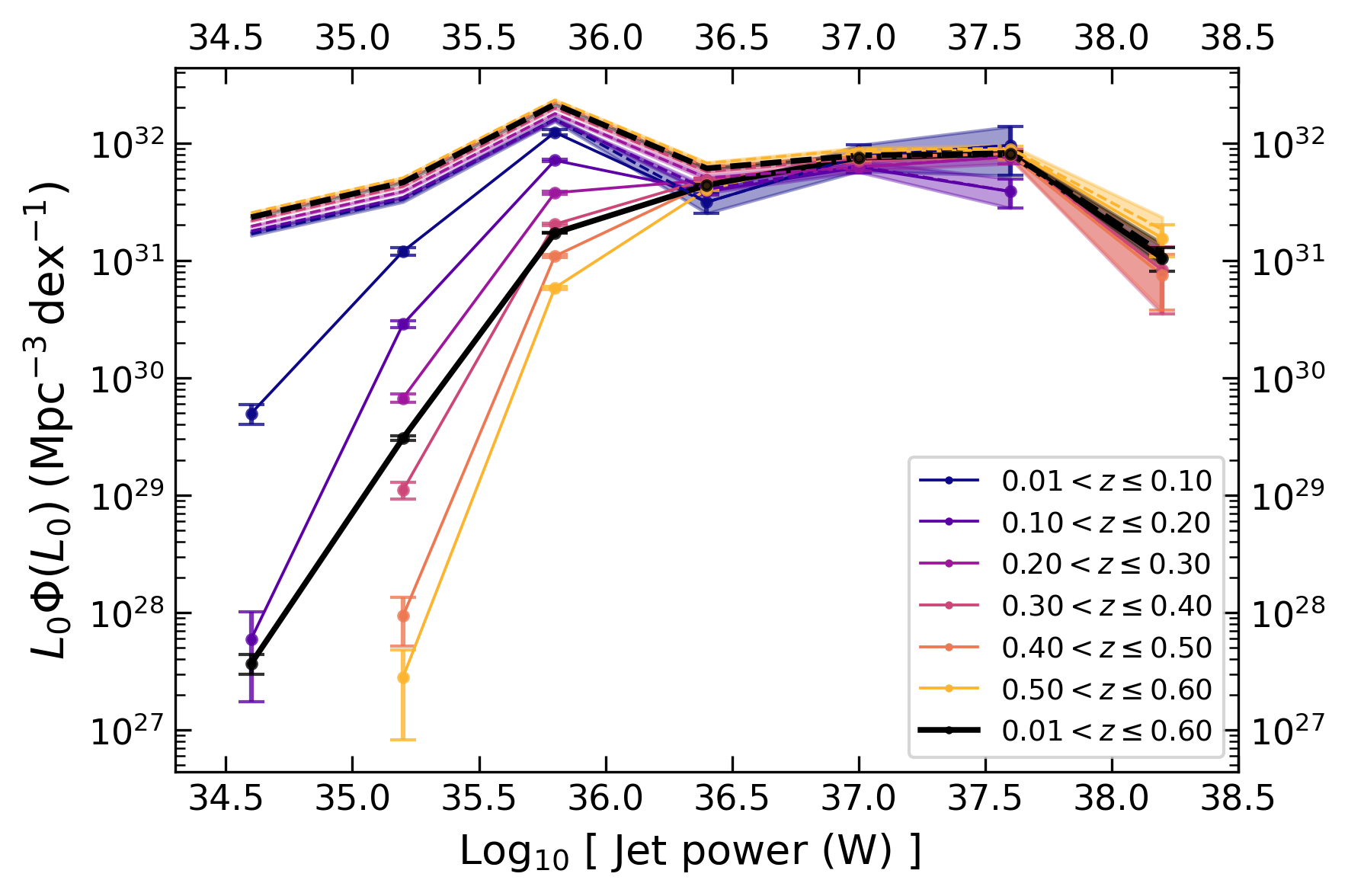}
\end{center}
\caption{Left panel: Jet power (kinetic luminosity) function of the OSS (solid lines and points with error bars) and the CSS (dashed lines, with errors represented as shaded areas) for different bins of redshift. The errors correspond to the Poissonian uncertainties. Right panel: Kinetic luminosity function multiplied by jet power, $L_0$, for the different bins and the resulting total for $z\leq 0.6$ (dark, thick lines).}
\label{jet_power}
\end{figure*}

Finally, Fig.~\ref{jet_power} shows the jet power function for the OSS and CSS (left) and this function multiplied by the jet power (right). This plot can be very useful from a theoretical perspective, e.g., we can use this result to estimate the amount of energy released by AGN in a cosmological context. We can estimate that the total injected energy by the complete sample would be $\sim 10^{56}\,{\rm J}$,\footnote{This can be directly derived from the simulation, by simply integrating the total energy injected by all the simulated sources at their respective observing time.} out of which the observable subset accounts for a 44.3\%. Furthermore, we have calculated the energy input by giant radio galaxies (those with projected sizes larger than 0.7 Mpc), which involve large volumes, and account for 14.4\% of the total, even if they represent a small fraction of the sample.

The right panel of Fig.~\ref{jet_power} shows the power function multiplied by power to give the energy input for different powers per comoving volume, as given by our OSS (solid) and CSS (dashed) samples. We show the result for the different redshift bins and the total (black, thick solid and dashed lines). For the OSS, we see that the energy contribution is dominated by sources at the high-power range of the distribution, in agreement with estimates shown in Figs.~A.5 in \citet{2019A&A...622A..12H} and Fig.~14 in \citet{2015ApJ...806...59T}. However, our simulated sample results in a less peaked distribution than the one seen in \citet{2019A&A...622A..12H}, with ours showing a wider plateau extending down to powers $\sim 10^{36}\,{\rm W}$. Interestingly, for the CSS we see that the distribution flattens \citep[as seen also for low mass hosts and/or low luminosities in][]{2015ApJ...806...59T}, indicating an equivalent contribution from both low and high power sources, and the peak of the contribution appears precisely at $\sim 10^{36}\,{\rm W}$.

\section{Discussion}\label{sec:disc}

\subsection{Limitations} \label{sec:limi}

In the previous section, we have presented the results produced by our simulations in comparison with the RLAGN sample in LoTSS. By enforcing a jet power spectrum that recovers the luminosity distribution, and choosing typical galactic environments in our model, we see that the main parameters controlling the resulting size distribution are the ratio of thermal to relativistic electron energy densities in the cocoon, $k(t)$, controlled by the jet mass load, and the activity period distribution. 

Among the limitations of our work, we find the way in which we have selected the function $k(t)$, aimed in principle to reproduce the mass-loading numbers derived from numerical simulations \citep[][see Paper~I]{2014MNRAS.441.1488P,2021MNRAS.500.1512A}, and slightly modified now to improve our global fits. Although the selected shape of $k(t)$ is reasonable (with stronger mass-load within the galaxy core followed by an extended region with a drop in the mass-load, still within the host), in future work, we intend to use machine-learning techniques to derive the best option for this function. A similar approach should also be followed for the maximum age (or activity period) distribution. Another aspect that limits our modeling is the simplification of the particle spectral distribution, for which a single injection event is considered, i.e., we do not include the effect of reacceleration mechanisms such as turbulence in the lobes.

In the case of compact sources ($\lesssim 25\,{\rm kpc}$), we should expect free-free absorption to reduce the received flux at low frequencies, thus reducing the number of observable sources in our simulation. Within this set, the radio lobes of sources with linear sizes below $\simeq\,10$~kpc, i.e., Gigahertz-Peaked Spectrum sources and Compact-Steep Spectrum sources, are known to be strongly affected by synchrotron self-absorption at low frequencies. They have their spectral turnover frequencies at the GHz and hundreds of MHz, respectively \citep[see, e.g.][for a review]{2021A&ARv..29....3O}. The effects of absorption have not been taken into account in our work and thus limit the comparison between our simulation and the observed sample for this range of source sizes. Absorption could push a number of them below the observability threshold, thus forcing the model to produce a larger number of sources in the complete sample. This could be relevant if we consider that low-power sources are much more numerous and are more likely to be observed with these linear sizes in comparison to powerful sources.

Although we impose a constant jet power along the activity period, and this power can be variable, the time-scales involved in jet evolution allow us to use a constant value that can be considered as a mean value of the injected energy per unit time during the active phase.

Another simplification of the model is the spherical symmetry imposed to the ambient medium, which locates all our galaxies at the center of clusters. However, this is relaxed by the fact that a large number of sources do not reach the ICM, which limits the symmetry demand to the host galaxy and its immediate environment (up to the point reached by the RG). 

From the perspective of the link to the observed population, the low number of sources at low and high luminosities plus absorption in the former introduce large uncertainties in the RG power spectrum that is imposed to our simulation. The powerful RGs imply a residual number of sources, but the low power end dominates the counts and a small change in the slope may imply a difference of tens of thousands of sources in the CSS. Furthermore, although we have not considered remnants in our simulated sample, different works agree in their rapid fading and their minor contribution to the whole sample \citep[e.g.,][]{2019A&A...622A..12H,2020MNRAS.496.1706S}.

Finally, we have not accounted for the jet emission (and Doppler boosting) in the source evolution, although we expect it to be negligible for the largest part of the observed sources at low frequencies and low redshifts, as those in the subsample of the LoTSS survey that we have used here. This contribution is to be taken into account to study high redshift populations, owing to both the redshifted emission having been generated at higher frequencies, for which the jets may dominate the emissivity, and also being boosted by relativistic beaming.

\subsection{The role of the activity period}

First of all, our model does not account for recurrent activity because we assume that all the RLAGN LoTSS are active now, and that lobe radio luminosity falls rapidly once the nuclear activity ceases \citep[see Paper~I and][]{2018MNRAS.475.2768H,2020MNRAS.496.1706S}. 

We have studied the resulting simulated populations for different values or distributions of the maximum activity times. The results are shown in Appendix~\ref{app1}. In Fig.~\ref{apendix1} we show the size distribution for a global maximum age of the sources, using 5, 50 and 500~Myr. Although a global maximum of 50~Myr gives the distribution peak close to the observed one, none of the options seems to approach the LoTSS profile. 

Figure~\ref{apendix2} shows the results for different distributions of maximum times: a uniform distribution in linear and logarithmic scales between 5 and 500~Myr, and a normal distribution peaked at 100~Myr with an amplitude of 50~Myr. We see that the uniform distribution in logarithmic scale is the one that gives the best approach to the observed one. Still, in Fig.~\ref{apendix4} we show the results for three different normal distributions, which all produce an excess of sources beyond the observed peak. Finally, in Fig.~\ref{apendix3} we show the result derived for different intervals of $t_{\rm max}$ to which we apply the logarithmically uniform distribution. After this analysis, we decided to use the one that gave the best results: a logarithmically uniform distribution between 10 and 200~Myr.

It is relevant to stress that the distribution of maximum ages that results in the best fit to the LoTSS sample ($\propto t_{\max}^{-1}$, equivalent to a uniform distribution in logarithmic scale) coincides with that derived by \citet{2020MNRAS.496.1706S} and \citet{2025MNRAS.537..343Q}. Interestingly, \citet{2025MNRAS.537..343Q} reach this result using a completely different approach, based on the modeling of a sample of 79 remnant sources with the dynamical code RAiSE \citep{2023MNRAS.518..945T}. The authors relate this lifetime function  to that expected from feedback-regulated accretion and galactic activity. Our result could point in this same direction for a sample of thousands of sources. In conclusion, observed samples can allow us to distinguish between different plausible scenarios.

\subsection{The role of mass-loading}

Although it is difficult to establish a direct relation between the $k$ parameter and the mass-load of the jet, we can try to do it by taking $k=u_{\rm T}/u_{\rm e}=\rho_{\rm T}/\rho_{\rm e}$ (i.e., the thermal to non-thermal rest-mass densities), assuming that the specific energy per unit mass is the same for both populations. With this assumption, we can study an example case. If we consider a leptonic jet with number density $1\,{\rm cm^{-3}}$ when its radius is 1~pc, and assume conical expansion from a radius of 1 to 100~pc, its number density at the end of this expansion would be $\sim 10^{-4}\,{\rm cm^{-3}}$ \citep[these can be typical values for jets with $L_{\rm 0}\,=\,10^{36}\,{\rm W}$, e.g.,][at $\simeq 1$~kpc]{2021MNRAS.500.1512A}. Adding a value of $k=10^3$, as reached by low power jets (see Paper~I), we can derive a proton density $\rho_{\rm p}\sim 10^{-26}\,{\rm kg/m^3}$. For a jet radius of $\sim 100\,{\rm pc}$, we then get $\rho_{\rm p}A_{\rm j} \sim 10^{32}\,{\rm kg\,kpc^{-1}}\simeq 50 \,{\rm M_\odot\,kpc^{-1}}$ (where $A_{\rm j}$ is the jet cross section), which is within the order of magnitude of the linear densities estimated in \citet{2002MNRAS.336.1161L} at the inner kiloparsecs and plotted in their Fig.~8. However, in that paper the authors consider ongoing mass-load with distance along 12~kpc, whereas we limit it to the inner three core radii ($\leq 6$~kpc). 

This direct relation between the different functionalities used for $k(t)$ and mass-load have allowed us to probe the role of this relevant process on the length distribution of the simulated sample. Figures~\ref{kevolucio} and \ref{apendixck} show the difference in the evolution of luminosity and the resulting distributions for the three different evaluated profiles of the mass-loading function. Figure~\ref{kevolucio} represents a clear example of the critical role that mass-loading can play in long term radio emissivity of radio galaxies, mainly for low power cases, where the variation in the calculated luminosity can be larger than an order of magnitude for different load profiles. These profiles are expected to depend on the host galaxy properties (namely, gas density and stellar population) and their proper characterization in analytical models is crucial to model the low power, dominant, AGN population. As we have seen, our model indicates that this population is possibly responsible for more than half of the energy injected by galactic activity into the host galaxies and/or the warm/hot intergalactic medium.

The dashed curve in Fig.~\ref{apendixck} (using a relatively low value of the constant $c_k$ in Eq.~\ref{k_paperI}) shows an excess of compact sources when applying the same function that we used in Paper~I (for a constant mass-load rate). However, this peak lies on the region (1-10~kpc) where the reliability of the comparison between our model and observational samples can be severely limited, as mentioned above, but we regard the effect of the mass-load as valid from a modeling point of view. In other words, a reduced mass-load allows low power (and typically smaller) sources to be brighter, which results in a relative increase of observed compact sources and a corresponding deficit of the large ones. In this case, we can derive conclusions by comparing the numbers obtained for large sources in the modeled sample with the observed one because they are not affected by absorption, and we indeed see a relative deficit in the simulation.

We see that an increase of the mass-load reduces this excess and, at the same time, shifts the peak of the distribution towards larger source sizes. At large scales, where we can compare the observed and simulated samples, we see that the peak of the LoTSS distribution is located at $\simeq 30$~kpc, whereas the increase of the mass-load shifts it to $\simeq 100$~kpc in the simulation. It thus appears that the power spectrum derived from the fit to the luminosity distribution produces an increase in the number of powerful sources to compensate the strong mass-load. As a result, we obtain this excess of larger sources. Furthermore, from Fig.~\ref{kevolucio}, we see that intermediate power sources show an increase of their radio luminosity beyond $\simeq 100$~kpc, which could make them preferentially observable at those sizes in cases around the detectability limit. This second option could be, nevertheless, limited by the drop in surface brightness when the lobes reach hundreds of kpc. 

If we split the mass-load rate in two regions, with Eq.~\ref{k_paperII}, the comparison between the dotted and solid lines in in Fig.~\ref{apendixck} show that the reduction of the entrainment rate in the second phase (between the core radius and three times this distance) results in a relaxation of the aforementioned need for powerful sources, which reduces the excess at large scales. At the same time, the high mass-load rate within the galactic core maintains the requirement of larger powers to produce a given luminosity. The consequence is the disappearance of the excess of compact sources and the reduction of the difference with the peak of the observed distribution ($\simeq 50-60$~kpc cf. $30$~kpc for the LoTSS sample). Therefore, a combination of high mass entrainment within the core and an ongoing, milder value farther downstream seems to produce an appropriate combination to explain the observed size distribution (at least at the scales where the comparison is meaningful).

The relative deficit of simulated sources between hundreds of kpc and Mpc scales observed in Fig.~\ref{histD} is probably caused by the selected mass-load function, as we justify in the following lines. This feature is redshift independent (see Fig.~\ref{Lvsz}) and is also visible in Fig.~\ref{LvsD}. The deficit is driven by the imposed power spectrum, and it is possibly caused by the large error associated to the relatively low number of powerful sources in the simulated sample. Our model seems to fit the high luminosity distribution in Fig.~\ref{histL} with jet powers that are lower than the ones required to generate a number of extended sources similar to the LoTSS sample. The introduction of Eq.~\ref{k_paperII}, limits the mass-load beyond the galactic core and thus favors that low power sources are brighter than they would be if we used Eq.~\ref{k_paperI} with a large value of base entrainment (see the comparison of the dotted and solid lines in Fig.~\ref{apendixck}). Altogether this means that we can reproduce the luminosity distribution with jets that have lower powers than we should use in order to obtain more large-scale sources. 

The approach between the compact observed and simulated samples seen in Fig.~\ref{apendixck} may be due to the fact that mass-load has the same effect as absorption, namely it reduces the luminosity of the sources. In this respect, it is possible that the value for mass-load that we use can be reduced when absorption is considered. At large scales, where absorption is negligible, the samples can be compared and thus used to calibrate the mass-load function, which, as we have shown, is a key factor in the understanding of the evolution of radio sources across the $P-D$ diagram. Other factors, such as the ambient density and galaxy size also play a significant role in the radio emissivity, as it has been repeatedly shown (see Paper~I and references therein), but we believe that these are already accounted for within the 
wide region of the parameter space that is used in our study.

It is clear that a proper definition of the mass-load function is crucial to reproduce accurately the observed populations. Figure~\ref{kevolucio} shows that mass-load can be fundamental for luminosity evolution for jet powers $\leq 10^{37}\,{\rm W}$, which involves the largest amount of sources (see Sect.~\ref{power_dist}). Therefore, omitting this ingredient can set limits to the interpretation of results obtained by analytical models for low power sources, and to any conclusion derived from them. Mass-loading could be, for instance, an extra factor to be added to the list given by \citet{2019A&A...622A..12H} when discussing the lack of agreement between the analytical model applied \citep{2018MNRAS.475.2768H} and the LoTSS sample. It could also be a relevant fact to be considered, even if indirectly, in the discussion of \citet{2020MNRAS.496.1706S} about the direct relation between the frequency of detection of remnant or restarted sources and the abundance of low power sources. Mass-load could rapidly reduce the emissivity of short lived, low power outflows, thus reducing their detectability and limiting our capability to estimate their abundance. We believe that our approach may help to overcome this limitation because our complete sample is built taking mass-loading into account. Future work should tackle these aspects in more detail. 

An approach to the properties of the mass-load function in jets would benefit from studies of stellar populations in AGN jet host galaxies, or determining jet velocity profiles \citep[\textit{alla} Laing \& Bridle, e.g.,][]{2014MNRAS.437.3405L}. Such investigation could facilitate a better characterization of mass-load and, thus, a better determination of radio source evolution through the $P-D$ plane. The implications that this may have on the study of AGN feedback and host galaxy evolution through cosmic ages are expected to be very relevant, judging from our results.

\subsection{Future prospects}

After validating and applying our model we have identified two fundamental ingredients of RG evolution and resulting simulations, namely the mass-loading and duration of the nuclear activity. By selecting the optimal distribution of these parameters among a few studied options, we have obtained an acceptable fit to the LoTSS RLAGN luminosity and projected lengths distributions.  

Model improvements in the directions indicated by its current limitations (Sect.~\ref{sec:limi}) should be undertaken. In particular, a devoted study of mass-load and activity periods should be performed, together with a revision of other important parameters in our results, such as, for instance, galactic and cluster pressure/density profiles. Furthermore, implementing the physical processes that would allow us to extend our work to higher redshifts (e.g., jet emission from boosted, aligned sources) will be crucial to study the relevance and role of galactic activity through cosmological ages.

\section{Summary and conclusions}\label{conclusions}

In this work, we have presented a methodological path to probe the populations of active galaxies at low redshift. Using the dynamical model presented in Paper~I we have simulated the evolution of hundreds of thousands of sources with different properties and environments and compared them to the LoTSS RLAGN sample. We show that the selection of a mass-load profile in jets has to be regarded as a key ingredient in the modelization of radio source evolution, mainly for the low power sources, which are, in addition, the most numerous and could account for more than half on the energy released by AGN into their hosting clusters and groups.

As a first step, we have applied our model to fix a RLAGN power spectrum that allows us to accurately fit the observed luminosity distribution. Then, we use this power spectrum along with probability distributions of other RG and host galaxy/cluster properties to produce a mock population. We performed this process five times in order to derive the number of total sources that there may be in the LoTSS field for LOFAR to detect the number in the RLAGN sample. Our results show that at least $\simeq 3.8\times10^5$ AGN with radio emission must exist in order to obtain 17045 observable sources with LOFAR. However, this number depends on the power distribution, mainly in the low power regime, which is uncertain due to the smaller number of observed low-luminosity sources, and it thus also depends on the exact role that absorption may have in hiding the population of compact sources. Moreover, from these numbers we estimate that a telescope with a sensitivity threshold of $0.01$~mJy would detect one order of magnitude more sources. 

According to our calculation, the RLAGN population in the LoTSS field could have injected $\sim 10^{56}$~J into their host galaxies and environments. $44\%$ of this energy would correspond to the observed sample.

Among other less critical parameters, we have identified the mass-load undergone by the jet within the host galaxy and the activity times as crucial parameters to reproduce the observed distributions and derive fundamental information, such as expected RLF at higher redshifts and power spectrum along the cosmological ages. Our best model successfully approaches the observed distributions using a mass-load profile that increases with distance through the galactic core and keeps increasing up to three times the core radius at a slower pace (mimicking a drop in stellar and gas densities) and a maximum activity time spread uniformly in a logarithmic scale between 10 and 200~Myr. This result agrees with previous independent work and points towards feedback-regulated activity time-scales \citep[see][]{2020MNRAS.496.1706S,2025MNRAS.537..343Q}. Future work should focus on improving the characterization of these relevant parameters, adding jet contribution, exploring ways to include galactic extinction, and eventually extending this work to larger redshifts.

\begin{acknowledgements}

The authors thank the referee of this paper, Stas Shabala, for constructive criticism and comments that have certainly helped us to improve this work. We also thank Martijn Oei for useful discussions. This work has been supported by the Astrophysics and High Energy Physics program supported by the Spanish Ministry of Science  and Generalitat Valenciana with funding from European Union NextGenerationEU (\texttt{PRTR-C17.I1}) through grant \texttt{ASFAE/2022/005}, by the Spanish Ministry of Science trough Grants \texttt{PID2022-136828NB-C43} and \texttt{PID2022-138855NB-C33}, and by the Generalitat Valenciana through grant \texttt{CIPROM/2022/49}.
\end{acknowledgements}

\bibliographystyle{aa.bst}
\bibliography{aa59768-26}

\begin{appendix}
\section{Mass-load and activity periods}
\label{app1}

\begin{figure}[htbp]
\begin{center}
\includegraphics[width=85mm]{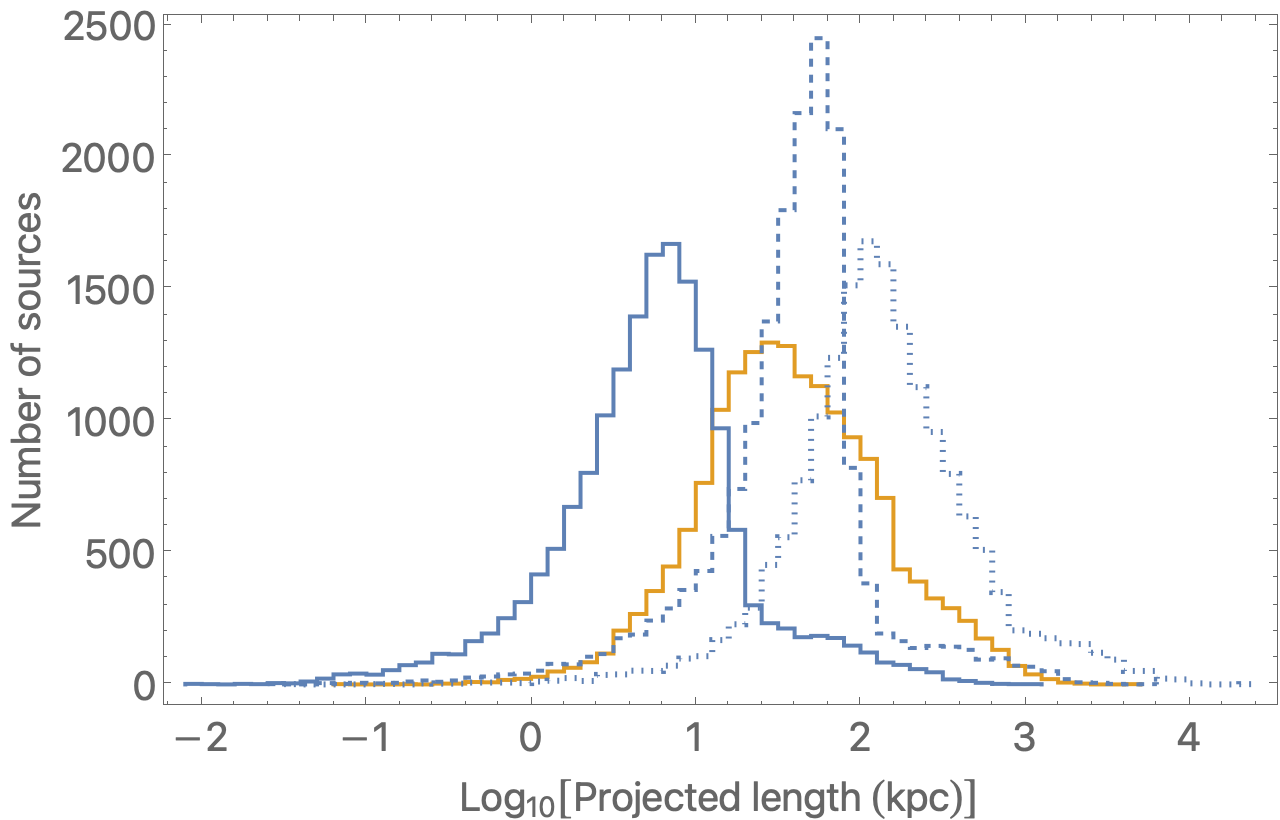}
\end{center}
\caption{Histograms of projected length of the observations (orange) and three simulations (blue) for different fixed values of $t_{\rm max}$: 5 Myr (solid), 50 Myr (dashed) and 500 Myr (dotted).}
\label{apendix1}
\end{figure}

\begin{figure}[htbp]
\begin{center}
\includegraphics[width=85mm]{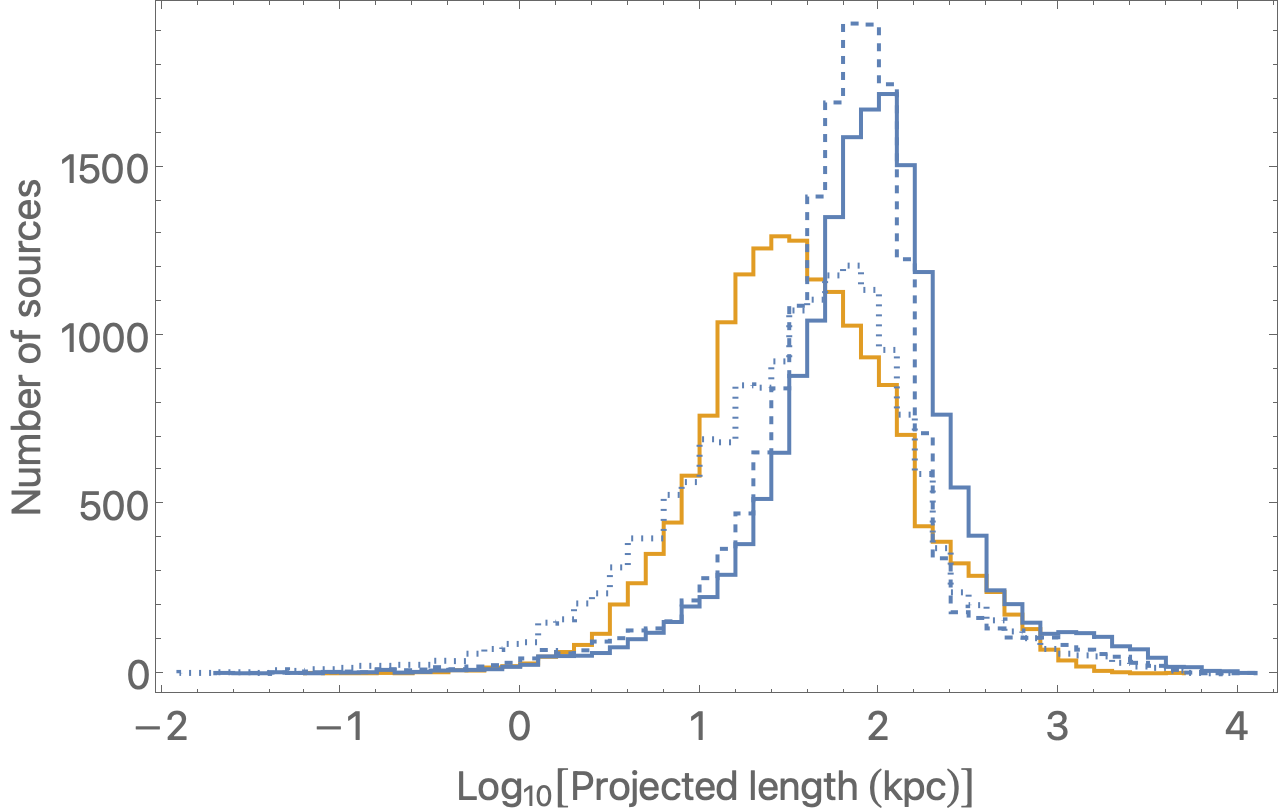}
\end{center}
\caption{Histograms of projected length of the observations (orange) and three simulations (blue) for different distributions of $t_{\rm max}$: uniform between 5 and 500 Myr (solid), normal centered in 100 Myr with standard deviation of 50 Myr and truncated for $>1$Myr (dashed), and uniform in logarithmic scale between 5 and 500 Myr (dotted).}
\label{apendix2}
\end{figure}

As explained in Sect.~\ref{param}, the intervals we apply for the randomization of most parameters are based on the results from Paper~I. However, we have had to adjust some parameters in order to obtain a distribution of projected lengths that approaches the observed one. These are 
the maximum jet activity time ($t_{\rm max}$) and the evolution of the ratio of thermal to magnetic energy densities, $k(t)$ (see Eq. (17) in Paper I), interpreted as driven by mass-load in our model. 
In this appendix, we present a selection of the tests we carried out by varying these parameters to justify the choice of the values that we use throughout the paper. In all the tests shown below, we first perform the optimization of the jet power distribution described in Appendix~\ref{app2} and then evaluate the full simulation, ensuring that the luminosity distribution matches the observed one in all cases. For this reason, we only show the distribution of projected lengths, which determines the quality of our fit, taking into account that we do not apply a condition in this case. In all the figures of the Appendix, all the parameters (with the exception of the one under analysis) are fixed or randomized as indicated in Sect.~\ref{method_description}.

\begin{figure}[htbp]
\begin{center}
\includegraphics[width=85mm]{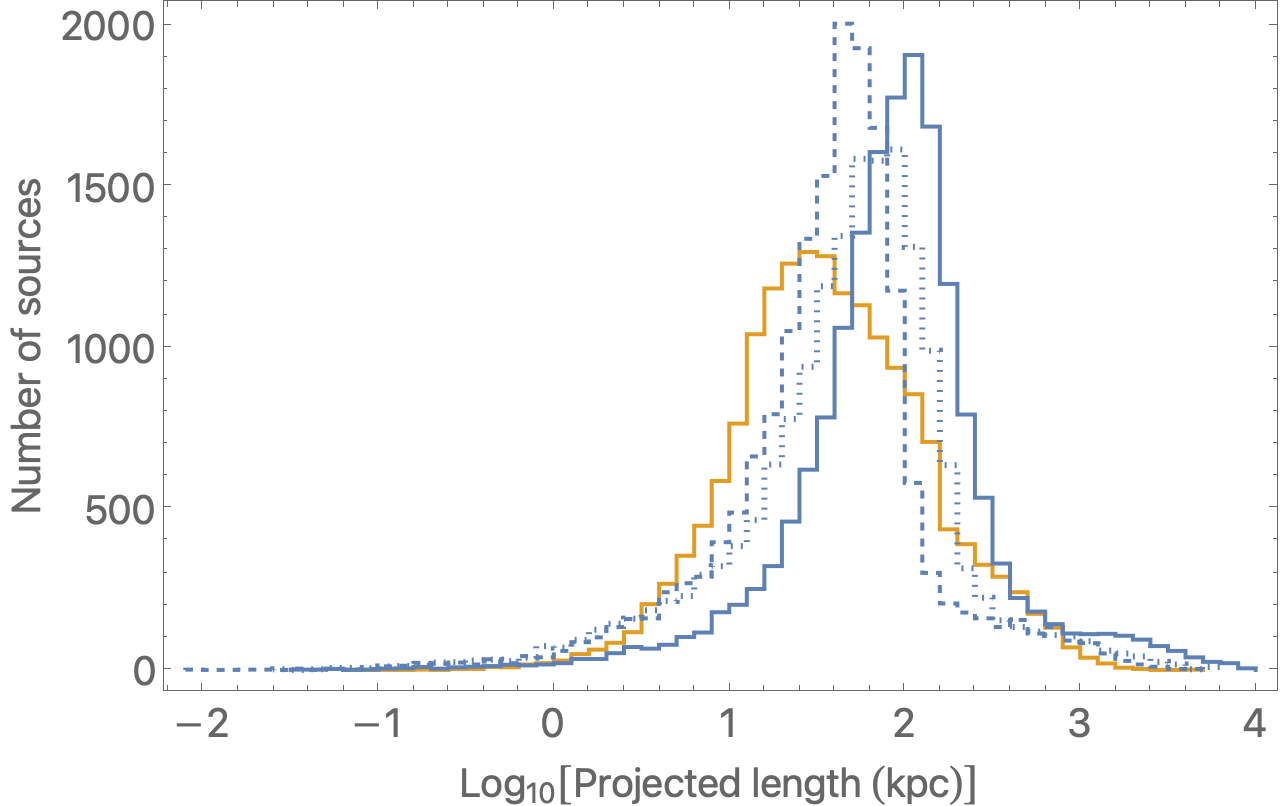}
\end{center}
\caption{Histograms of projected length of the observations (orange) and three simulations (blue) for three different normal distributions of $t_{\rm max}$ truncated at $>1$Myr: centred at 200Myr with standard deviation of 100Myr (solid), centred at 50Myr with standard deviation of 20Myr (dashed) and centred at 30Myr with standard deviation of 100Myr (dotted).}
\label{apendix4}
\end{figure}

\begin{figure}[htbp]
\begin{center}
\includegraphics[width=85mm]{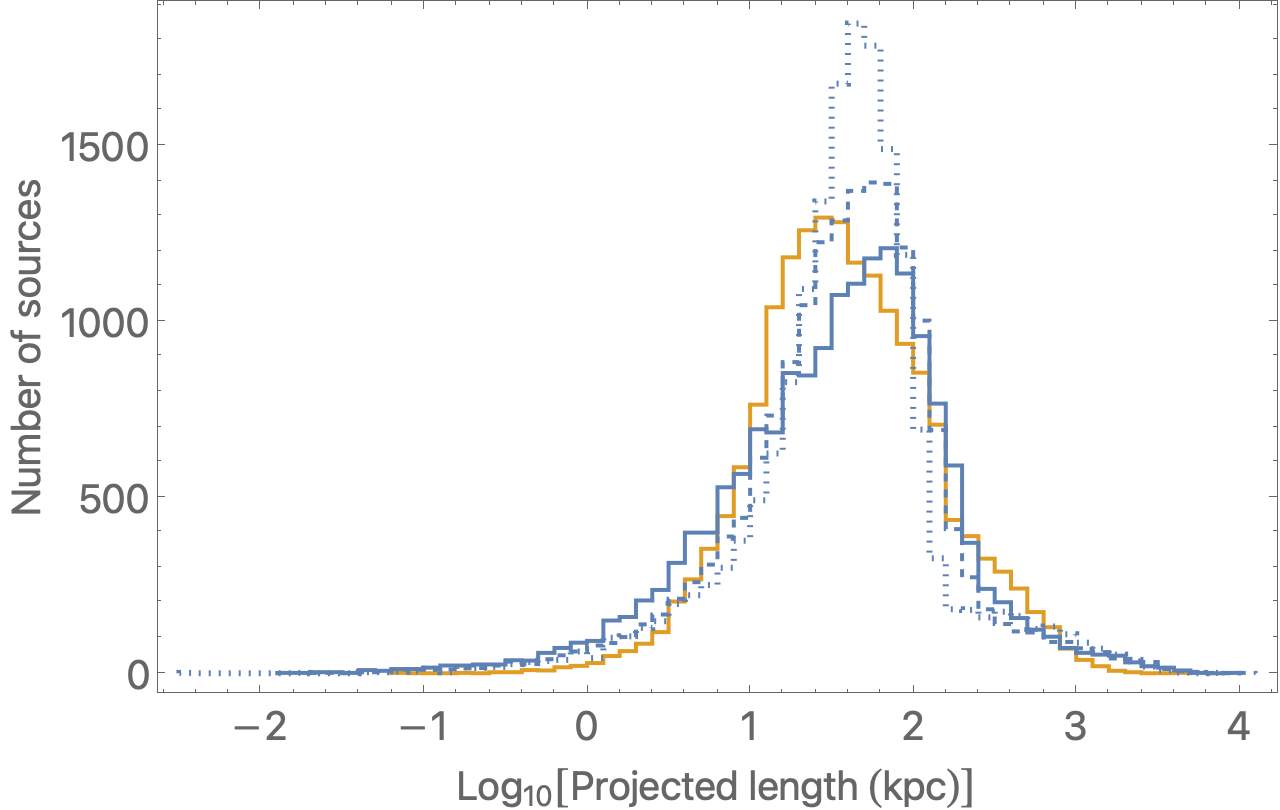}
\end{center}
\caption{Histograms of projected length of the observations (orange) and three simulations (blue) with $\log_{10}(t_{\rm max})$ distributed uniformly between different limits: 5-500 Myr (solid), 10-200 Myr (dashed) and 20-100 Myr (dotted).}
\label{apendix3}
\end{figure}
It is important to note that the comparisons in this appendix were made visually, since we have not still incorporated an optimization method for this distribution, in contrast to the radio luminosity. Therefore, this selection should be understood as a qualitative approach. A more rigorous method could be implemented using machine learning techniques to optimize the entire parameter set, which we leave for future work.

In Paper~I we concluded that the maximum activity time, $t_{\rm max}$, should take values below $500$ Myr (see Fig.~7 in that paper). In Fig.~\ref{apendix1} we show the size distributions corresponding to three fixed values of $t_{\rm max}$, along with the observed distribution. Since none of them seems to match, we decided to apply an a priori distribution of values and take random values within that distribution. In Fig.~\ref{apendix2} we show the histograms corresponding to three different cases: a uniform in linear and logarithmic scales (i.e., $\log_{10}(t_{\rm max})$ is uniformly distributed), and a normal. We can see that the logarithmically uniform distribution provides the best results. We have also tested three other normal distributions, which are shown in Fig. \ref{apendix4}, and reached the conclusion that none of them matches the LoTSS projected length distribution. Finally, in Fig.~\ref{apendix3} we present different tests for the logarithmically uniform  distribution by varying the randomization interval. Out of the different tests that we run, we selected a distribution of $t_{\rm max}$ in which $\log_{10}(t_{\rm max})$ is uniformly randomized between $t_{\rm max}=10$ and $200$~Myr, because it is the one that best reproduces the observed distribution. 

\begin{figure}[htbp]
\begin{center}
\includegraphics[width=85mm]{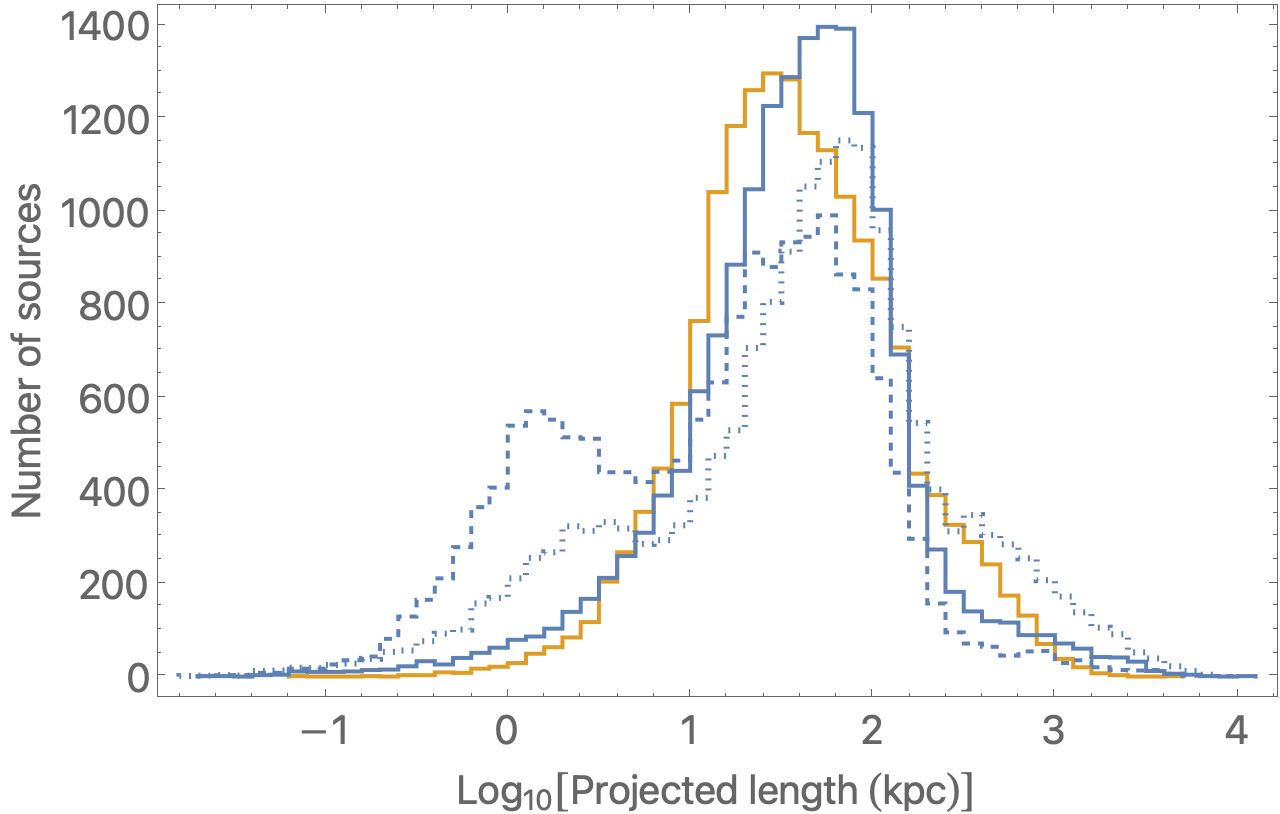}
\end{center}
\caption{Histograms of projected length of the observations (orange) and three simulations (blue) for different functions of $k(t)$: Eq. \eqref{k_paperI} with $c_k=10^6W^{1/2} s^{-1}$ (dashed), Eq. \eqref{k_paperI} with $c_k=10^7 W^{1/2} s^{-1}$ (dotted) and Eq. \eqref{k_paperII} with $c{k,1}=10^7 W^{1/2}s^{-1}$ and $c_{k,2}=10^6 W^{1/2}s^{-1}$ (solid).}
\label{apendixck}
\end{figure}

In Fig.~\ref{apendixck} we show a comparison of the projected length histograms obtained from three different simulations corresponding to three different expressions for $k(t)$: Eq.~\eqref{k_paperI} with $c_k=10^6 \,{\rm W^{1/2} s^{-1}}$ (the function used in Paper I), Eq.~\eqref{k_paperI} with $c_k=10^7\,{\rm W^{1/2} s^{-1}}$, and Eq.~\eqref{k_paperII}. We also show the distribution given by the LoTSS sample for comparison. We can see that the function adopted in Paper I with both values of $c_k$ produces an excess of compact sources (see Sect.~\ref{sec:disc}). In contrast, the change of slope in the evolution of $k(t)$ results in a closer approximation of the observations.

\section{Generation of the jet power distribution}
\label{app2}

In order to derive a jet power distribution that allows us to reproduce the observed luminosity distribution, we applied the method that is described in this Appendix. The power distribution is modelled as a piecewise function defined by a set of parameters $q_i$, as explained in Sect.~\ref{power_dist}. By applying the semi-analytical model of jet evolution, we can derive a simulated population for each parameter set. We then define a function that, for a given set of $q_i$, measures the difference between the luminosity distribution of the simulated and observational samples. Finally, we apply the Nelder-Mead method to find the values of $q_i$ that minimize this function.

A direct application of this method would require a large amount of time, as each simulation would need to be repeated thousands of times. Taking into account that each one takes between 5 and 10 hours (using 16 cores of 2.7 GHz in a local server), this is prohibitive with the current approach. Therefore, we implement an approximate method in order to accelerate the process, which consists on estimating the simulated luminosity distribution by applying relations of theory of probability. 

For simplicity, in subsequent calculations we define the variables $ x = \log_{10}(L_0) $ and $ y = \log_{10}(L_\nu) $ and use these variables instead of $ L_0 $ (power) and $ L_\nu $ (luminosity). 

\begin{figure}[htbp]
\begin{center}
\includegraphics[width=85mm]{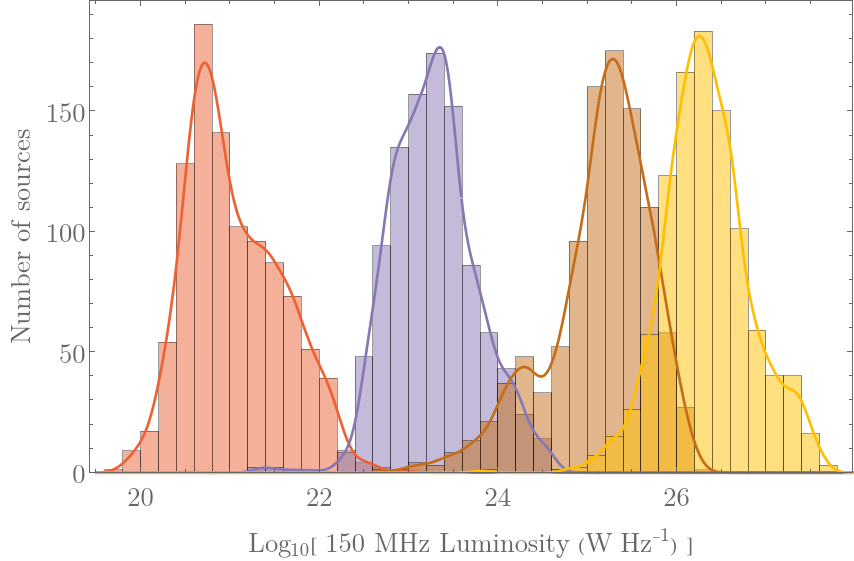}
\end{center}
\caption{Histograms of luminosity of simulated samples of 1000 sources for different fixed jet powers: $10^{35}W$ (red), $10^{36}W$ (purple), $10^{37}W$ (brown) and $10^{38}W$ (yellow).
The solid lines are the fits to the probability distribution that is then used to produce a RLAGN power spectrum (see the text).}
\label{distLL0}
\end{figure}

The relation between the jet power distribution, $P(x)$, and the luminosity distribution before applying the flux limit, $P(y)$, 
is given by 
\be\label{dist_lum_aprox}
{P}(y) = \int P(y | x) P(x) \, dx\, ,
\ee
where $P(y | x)$ is the conditioned distribution of luminosities for a fixed jet power. We estimate this function by running short simulations of 1000 sources for different jet powers ($x=34.5, 34.6, ... 38$) and save the obtained luminosity histograms.\footnote{We also performed some tests with smaller intervals of the $x$ parameter and obtained similar results, which validated the interval selection.} Then we approximate $P(y | x)$ by fitting the histogram corresponding to the closest jet power (we do so by using the function "SmoothKernelDistribution" of Mathematica, which returns a distribution from a data list, and "PDF", which gives the probability function of a distribution). Figure~\ref{distLL0} shows the results for different jet powers together with the fitted probability distributions of emitted radio luminosity at $150~{\rm MHz}$. 

Using this approximation and Eq.~\eqref{dist_lum_aprox} we can obtain an estimate of $P(y)$ for any set of $q_i$. In addition, with the aim to maximize the efficiency of the process, we substitute the integral in \eqref{dist_lum_aprox} by the Simpson's rule at order 100, as done in the integration of the luminosity (see Sect. \ref{method_description}). 

Finally, to obtain the luminosity distribution after applying the flux limit, we proceed as follows.\footnote{We do not consider the surface brightness limit in this approach, which is explained and justified at the end of the section.} As seen in Sect.~\ref{ambient}, the redshift of the galaxies (for $z<0.6$) is assumed to be distributed as $ P(z) \sim z^2$. Assuming $ L_\nu $ and $ z $ are independent (which is a reasonable assumption at least for low redshifts), the RG distribution in the local Universe can be expressed as $ P(y) P(z) $. And, after applying the flux limit, it becomes $ P(y) P(z) \theta(S(z,y) - 0.5 \, \text{mJy}) $, where $ \theta $ is the Heaviside function and the flux $S$ is given by Eq.~\eqref{flux}. This expression requires choosing a value for the spectral index $\alpha$. We do not have this information from the observations, so we fix a typical value of $0.75$ (we analyze the dependence of the results on this parameter below).

Thus, the marginal distribution of $ y $ after applying the flux limit (and subject to normalization) is
\bea \label{distLF}
P(y | S > 0.5 \, \text{mJy}) &\propto& \int_0^{0.6} P(y)\, P(z)\, \theta(S(z,y) - 0.5 \, \text{mJy}) \,dz \propto \nonumber \\
&\propto& P(y) \,\text{Min}(z_{\text{max}}(y), 0.6)^3,
\eea
where $ z_{\text{max}}(y) $ is the maximum $ z $ at which a galaxy with luminosity $ 10^y $ can be observed, which can be obtained numerically from the implicit equation $ S(z_{\text{max}}, y) = 0.5 \, \text{mJy} $. 

We then compare the function $\eqref{distLF}$ with the luminosity distribution of the observations, which we define as $\tilde P(y | S > 0.5 \, \text{mJy})$. The latter can be obtained directly by fitting the data histogram from the LoTSS sample \citep[Fig.~\ref{histL} and][]{2019A&A...622A..12H}. We define the following function to measure the difference between both distributions for each set of parameters $q_i$:
\be
\small{\Delta P (q_1, \dots, q_7) = \sum_{y=\boldsymbol{Y}} \left(\tilde P(y | S > 0.5 \, \text{mJy}) - P(y | S > 0.5 \, \text{mJy})\right)^2\, ,}
\ee
where $\boldsymbol{Y}=[y_{\rm min}, y_{\rm min}+0.1, ... y_{\rm max}]$, being $y_{\rm min}$ and $y_{\rm max}$ the lowest and highest luminosity of the observed sample. Finally, we apply the Nelder-Mead method to obtain the $ q_i $ parameters that minimize this function. 

\begin{figure}[htbp]
\begin{center}
\includegraphics[width=85mm]{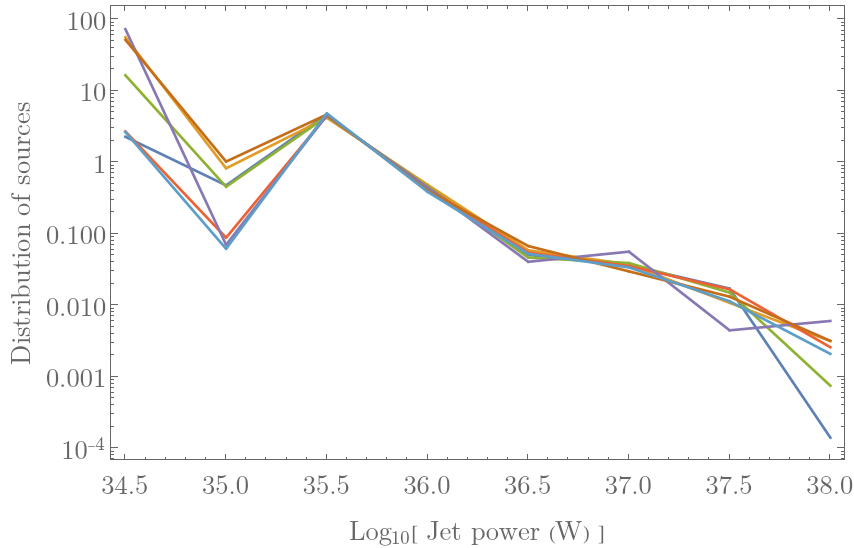}
\end{center}
\caption{Jet power distributions for the sets of parameters $q_i$ obtained in different tests.}
\label{distpot2}
\end{figure}

We repeated this process a total of seven times, and obtained different results because $P(y | x)$ changes slightly with the parameter randomization. In Table~\ref{table_qi} we show the results of the tests, and Fig.~\ref{distpot2} shows their corresponding jet power distributions. We then calculated the mean and standard deviation for each coefficient, and finally obtained the values shown in Eq.~\eqref{paramqi}. We also performed tests expanding the upper and lower limits of the jet power, but the results present higher errors because of the lack of observational data in those extremes. Figure~\ref{distpot} shows the power distribution $P(x)$ for the final set of parameters.

\begin{table}[h!]
\centering
\caption{Sets of optimized parameters $q_i$ obtained through seven different tests.\tablefootmark{a}}
\begin{tabular}{r r r r r r r}
$q_1$ & $q_2$ & $q_3$ & $q_4$ & $q_5$ & $q_6$ & $q_7$ \\
\hline
$-1.36$ & 1.99 & $-2.12$ & $-1.81$ & $-0.29$ & $-0.65$ & $-4.17$ \\
$-3.67$ & 1.42 & $-1.87$ & $-1.83$ & $-0.40$ & $-1.08$ & $-1.07$ \\
$-3.12$ & 1.97 & $-1.96$ & $-1.99$ & $-0.15$ & $-0.83$ & $-2.60$ \\
$-2.97$ & 3.41 & $-2.02$ & $-1.81$ & $-0.37$ & $-0.65$ & $-1.62$ \\
$-6.04$ & 3.62 & $-2.02$ & $-2.08$ & 0.28 & $-2.20$ & 0.26 \\
$-3.41$ & 1.32 & $-2.13$ & $-1.55$ & $-0.71$ & $-0.71$ & $-1.24$ \\
$-3.27$ & 3.79 & $-2.10$ & $-1.74$ & $-0.38$ & $-0.95$ & $-1.47$ \\
\end{tabular}
\tablefoot{\tablefoottext{a}{The values are normalized in the interval where they are more similar, between 35.5 and 37.5, for a better comparison.}}
\label{table_qi}
\end{table}

\begin{figure}[htbp]
\begin{center}
\includegraphics[width=85mm]{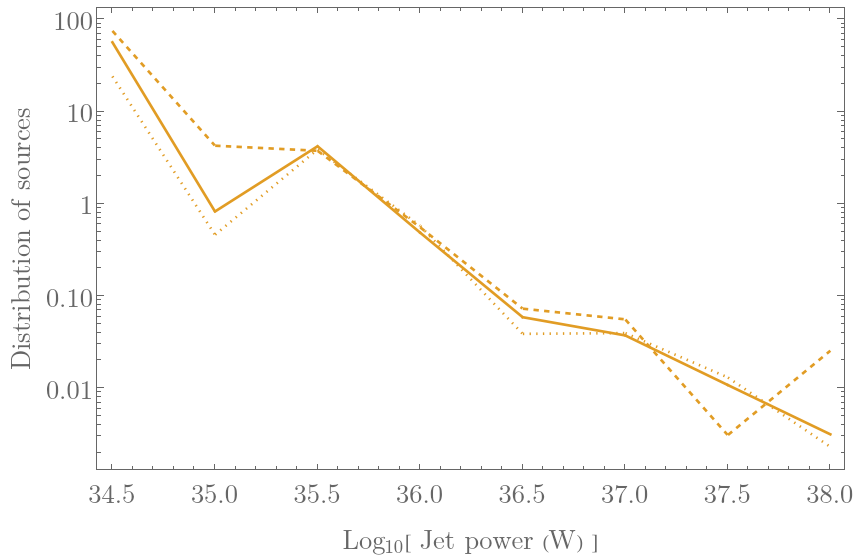}
\end{center}
\caption{Jet power distributions for one of the tests (second line in Table \ref{table_qi}) assuming three different values of $\alpha$: 0.75 (continuous), 1 (dashed) and 0.5 (dotted). They are normalized in the interval where they are more similar, between 35.5 and 37.5, for a better comparison.}
\label{distpotalpha}
\end{figure}

As mentioned above, in order to carry out this process we had to assume a fixed value of the spectral index, $\alpha$. To analyze the dependence of the result on this value, we show in Fig.~\ref{distpotalpha} the distributions obtained for one of the seven tests by varying the value of $\alpha$. We can see that in the central region of the jet power interval (where more sources are observed and the result is more reliable) the slopes of the distribution are similar. The differences found at the extremes
are comparable to those obtained for the different tests (see Fig. \ref{distpot2}). We therefore conclude that the uncertainty introduced by fixing a value of $\alpha$ will not significantly increase the error already considered previously.

\begin{figure}[htbp]
\begin{center}
\includegraphics[width=85mm]{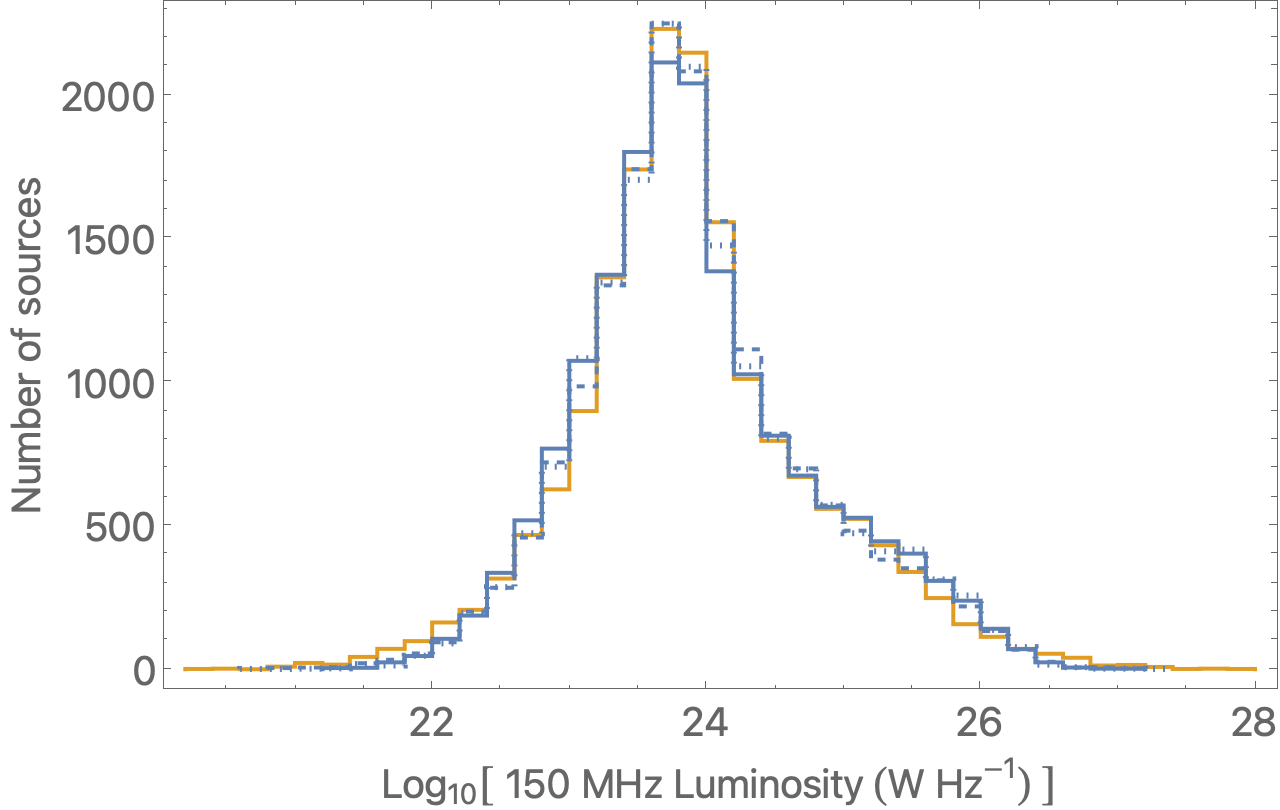}
\end{center}
\caption{Histograms of luminosity of the observations (orange) and three simulations (blue) for three of the seven sets of $q_i$ obtained.} 
\label{apendixLum}
\end{figure}

Finally, as a test of this optimization method, we plot the comparison of the histogram of source luminosity between the observations and three simulations in Fig.~\ref{apendixLum}, corresponding to three of the seven sets of $q_i$ that we obtained. We can see that in all cases the simulated histogram fits almost perfectly that resulting from the observations. We conclude, therefore, that the errors in the parameters $q_i$ are not significantly affecting the optimization of the luminosity distribution.

\begin{figure}[htbp]
\begin{center}
\includegraphics[width=85mm]{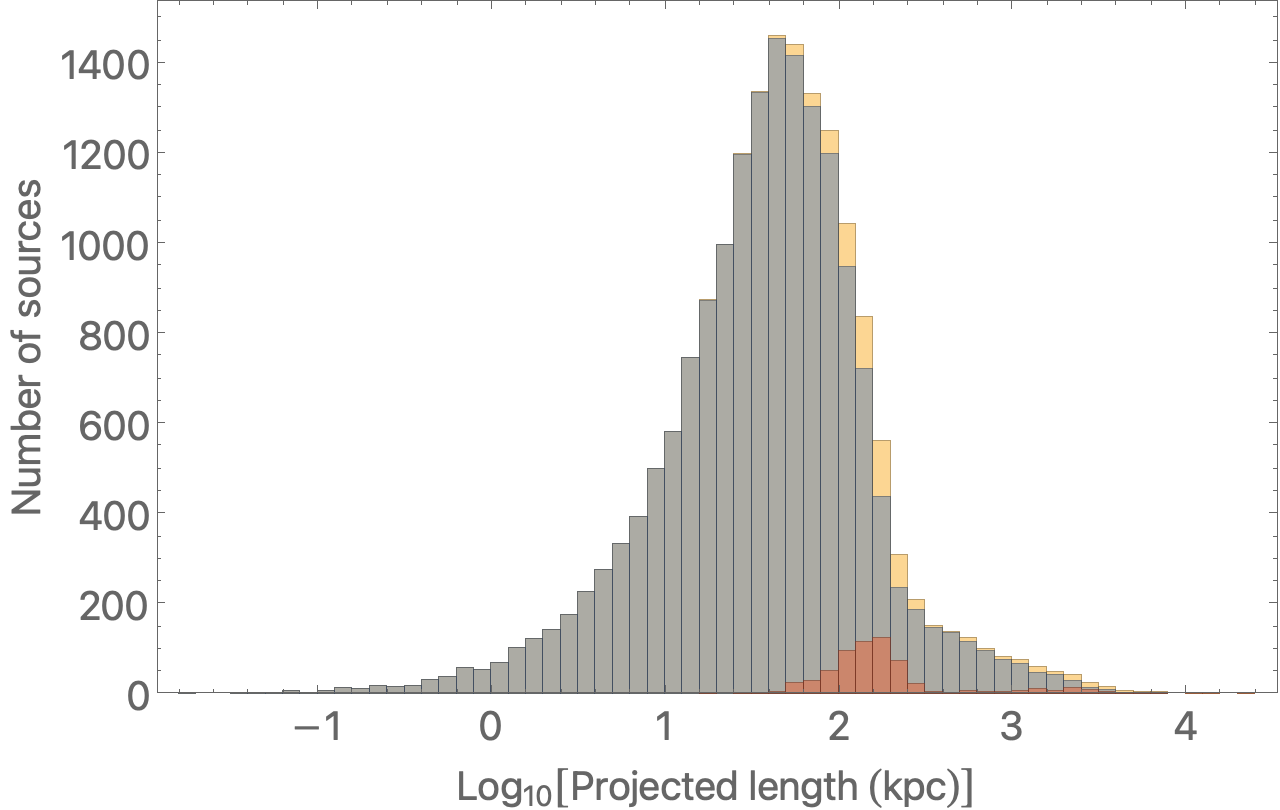}
\includegraphics[width=85mm]{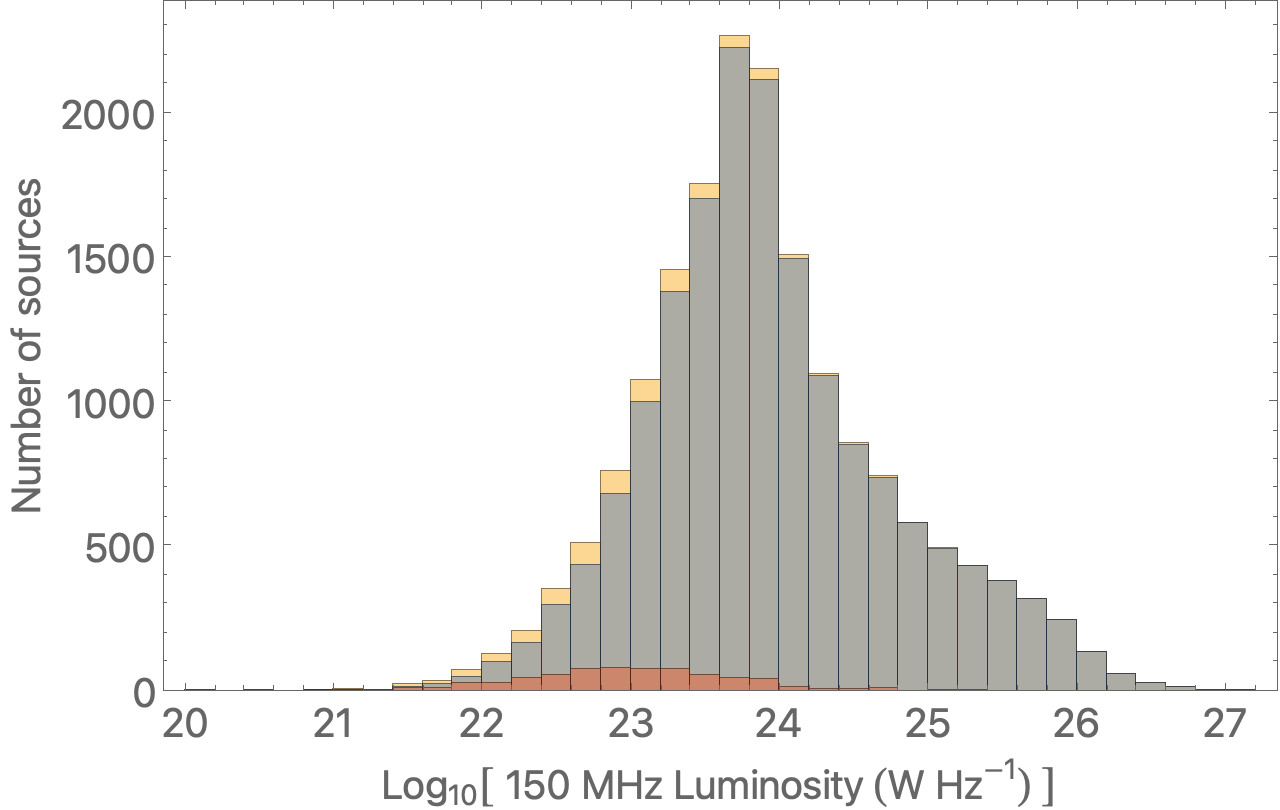}
\includegraphics[width=85mm]{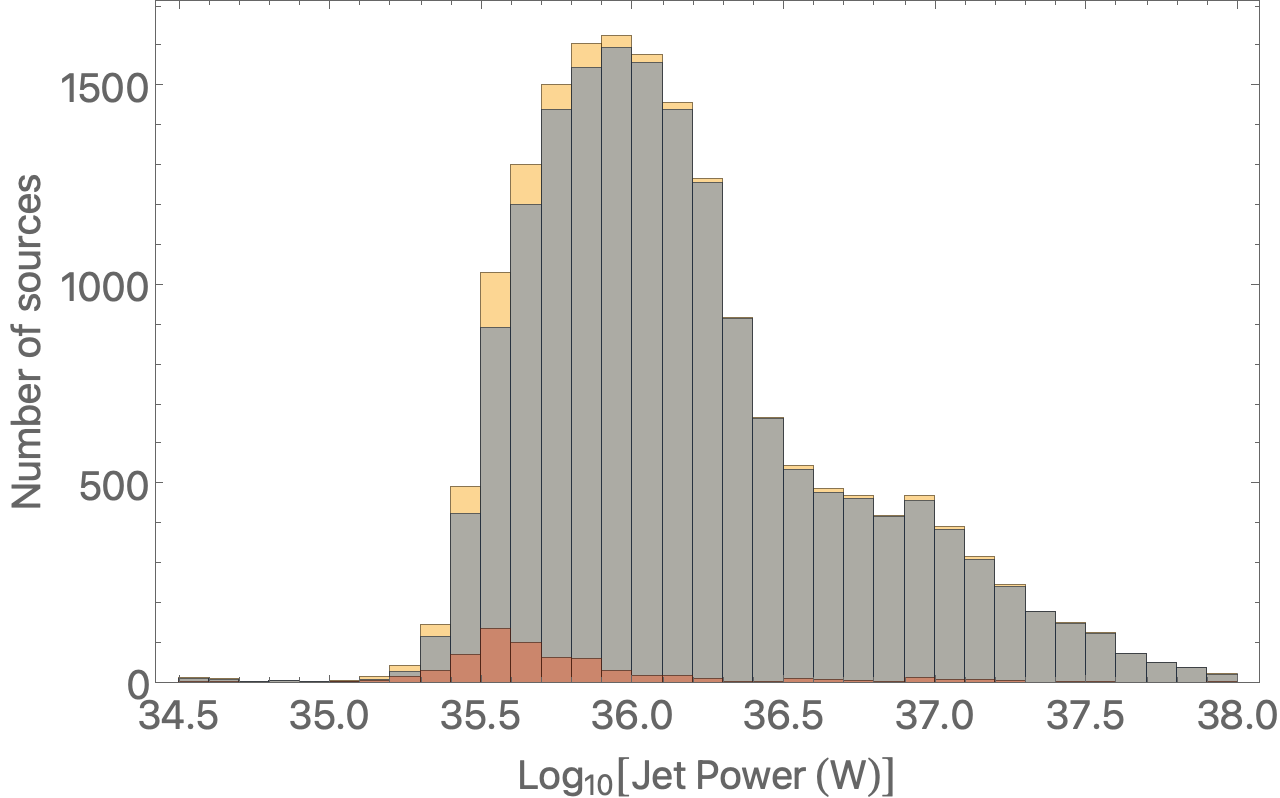}
\end{center}
\caption{Histograms of projected length (top), luminosity (centre) and power (bottom) of the OSS (gray), the set of sources that fulfil the total flux limit condition but not so the surface brightness one (red), and the union of both sets (the latter added to the OSS, orange).}
\label{histDcompapp}
\end{figure}

\begin{figure}[htbp]
\begin{center}
\includegraphics[width=85mm]{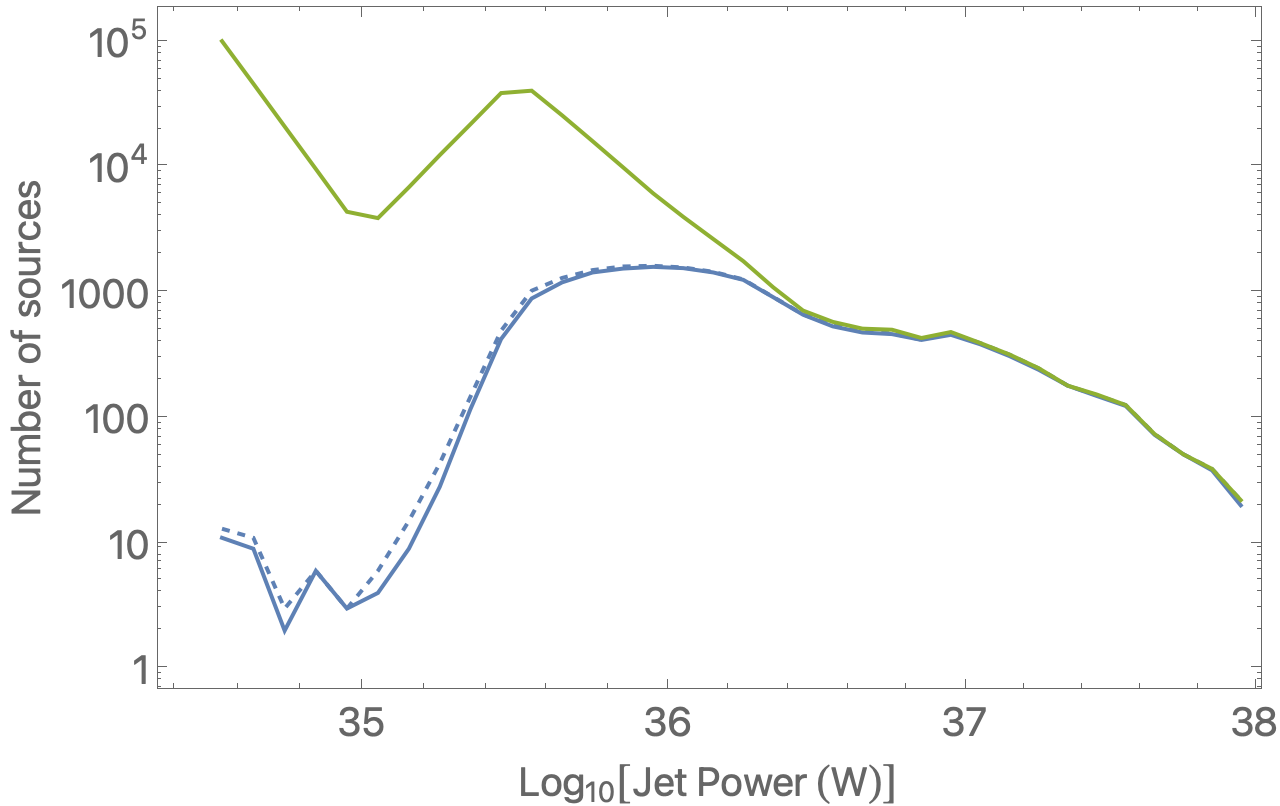}
\end{center}
\caption{Power distribution of the CSS (green), the OSS adding the sources that only pass the total flux filter (dashed blue), and the OSS with the sources that pass both thresholds (solid blue). The lines are formed by joining points taken from the heights of the columns of histograms produced with power bin widths ($\Delta (\log L_0)$) of 0.1.}
\label{histL0compapp}
\end{figure}

The steps described in the previous paragraphs omit the application of the surface brightness limit to our simulated sources in the determination of the power distribution. To justify this, we have to recall that the use of Eqs.~\ref{dist_lum_aprox} and \ref{distLF} implies the assumption that there is a dependence between jet power and luminosity, that the observability is only limited by the total flux, and that the source luminosity is fairly independent of its size. Introducing the surface brightness limit in this approach would break this assumption, because it adds a dependence on the size of the source. This would make it impossible to reduce the problem to the simple equations that have allowed us to implement the Nelder-Mead method. Therefore, we evaluated how relevant this second threshold could be for our estimates. If we omit this second threshold in the analysis and apply the power distribution derived, we see that the number of sources that fulfill the total flux limit but not so the surface brightness is 645 (3.8\%). Their size, luminosity and power distributions are shown in the Fig.~\ref{histDcompapp}. We see that the surface brightness filter affects sources with sizes above 50~kpc, low luminosity ($\leq 10^{24}\,{\rm W\,Hz^{-1}}$) and low power (below $10^{36.5}\,{\rm W}$). Taking into account that this power range is well populated in the power spectrum distribution, we do not expect a significant difference in the derived exponents. Figure~\ref{histL0compapp} shows the power distributions derived for the observable sources (solid, blue line), the observable sources plus those 645 below the surface brightness limit (dashed, blue line), and the power spectrum derived using the method described in this appendix (green). We see that the blue solid line deviates only slightly from the total distribution for large powers and that the surface brightness threshold mainly affects low powers. Moreover, Fig.~\ref{apendixLum} shows that, even considering this approximation, we have achieved our aim of finding a jet power distribution that reproduces the luminosities of the observations.

We think this analysis reinforces the assumptions made to derive the power distribution of the radio sources in the sample, at least in what respects the surface brightness limitation. In conclusion, this method allowed us to derive an estimation of the jet power distribution for which our model best matches the observed luminosity counts, and this is used to randomize the jet power values in the simulation.

\end{appendix}

\end{document}